# Numerical modelling of the proposed WFIRST-AFTA coronagraphs and their predicted performances

**John Krist, Bijan Nemati, Bertrand Mennesson**

Jet Propulsion Laboratory, California Institute of Technology, 4800 Oak Grove Drive, Pasadena, CA, USA, 91109

**Abstract**. The WFIRST-AFTA 2.4 m telescope will provide in the next decade the opportunity to host a coronagraph for the imaging and spectroscopy of planets and disks. The telescope, however, is not ideal, given its obscured aperture. Only recently have coronagraph designs been thoroughly investigated that can efficiently work with this configuration. Three coronagraph designs, the hybrid Lyot, the shaped pupil, and the phase-induced amplitude-apodization complex mask coronagraph (PIAA-CMC) have been selected for further development by the AFTA project. Real-world testbed demonstrations of these have just begun, so for now the most reliable means of evaluating their potential performance comes from numerical modeling incorporating diffraction propagation, realistic system models, and simulated wavefront sensing and control. Here we present the methods of performance evaluation and results for the current coronagraph designs.

**Address all correspondence to:** John Krist, Jet Propulsion Laboratory, 4800 Oak Grove Drive, Pasadena, CA, USA, 91109: Tel: +1 818-393-1194; E-mail: john.krist@jpl.nasa.gov

## 1 Introduction

NASA was given two space-qualified 2.4 m optical telescope assemblies that were spares from an Earth reconnaissance program. One of these has been designated the Astrophysics Focused Telescope Asset (AFTA)[1] for use by the NASA astrophysics program. The current plan, still under formulation, is to use it for the Wide Field Infrared Survey Telescope (WFIRST), a space mission to investigate dark energy and observe gravitational microlensing events caused by extrasolar planets. It has a proposed mid-2020's launch. WFIRST was rated as the top priority in the National Research Council's decadal survey of proposed astronomical projects for the 2010 – 2020 timeframe. A coronagraph for imaging and characterizing extrasolar planets and circumstellar disks has been proposed as an add-on to WFIRST-AFTA. It would be a separate instrument that picks off a small portion of the WFIRST field. Light from a star would be suppressed by a series of masks allowing a faint planet to be measured by a camera or integral field spectrograph (IFS).

The past decade of space coronagraph development has focused on potential future missions where it was assumed the telescope was specifically built for high contrast imaging, namely an off-axis design lacking any obscurations. AFTA, however, imposes an already-built telescope that was not designed for astronomy, let alone for coronagraphic imaging. It is an on-axis telescope with a central obscuration due to the secondary mirror and six fairly thick support struts (Fig. 1). This complex obscuration pattern significantly affects the design and performance of the coronagraph, and most current techniques require modifications to handle it.

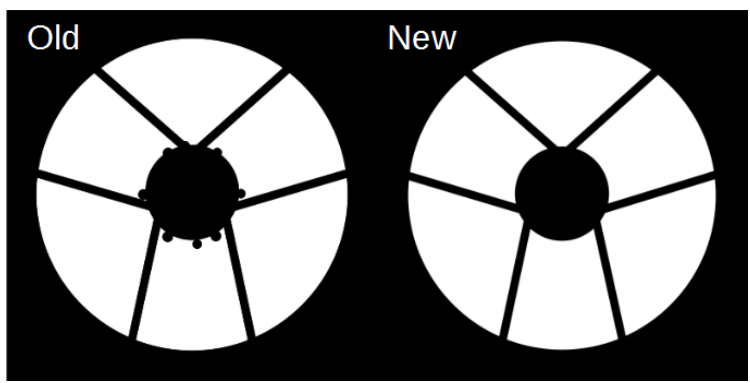

Fig. 1 - The AFTA telescope obscuration pattern. The "old" pattern on the left includes metrology markers along the edge of the secondary obscuration. Because the primary and secondary mirrors will be refigured and recoated, these will be removed so that the pupil will look like the "new" pattern on the right. All of the coronagraph designs presented here are for the "old" pattern except for the revised PIAACMC, which uses the new one (the markers do not significantly affect the performance either way).

A variety of configurations were proposed for the AFTA coronagraph: the hybrid Lyot coronagraph (HLC)[2], phase-induced amplitude apodization with complex mask coronagraph (PIAACMC)[3], shaped pupil coronagraph (SPC)[4], and a combination of a vector vortex coronagraph (VVC)[5], shaped pupil mask, and obscuration apodization using deformable mirrors (Active Compensation of Aperture Discontinuities, or ACAD)[6]. Two varieties of interferometric coronagraphs were also considered: a DAVINCI visible nuller coronagraph (VNC)[7] and a phase occulting VNC (POVNC)[8]. All of these techniques are still technologically immature to some degree, along with the wavefront sensing and control methods that will be required to compensate for optical aberrations and pointing errors. None of these have yet been demonstrated on testbeds with obscured pupils with the performance necessary in an AFTA flight instrument (some only exist in conceptual form). In order to meet the goal of deciding in the next couple of years if a coronagraph is a valid additional instrument for AFTA, any design would have to undergo accelerated development and hardware demonstrations.

Given the resource constraints of time, funding, and facilities, not all of these techniques could be supported, so an informed downselect was necessary. NASA formed the AFTA Coronagraph Working Group (ACWG) in 2013 to undertake an intensive study of their predicted performances and likelihoods of hardware flight readiness in the required timescale. Key to this was numerical modelling of each technique in a simulated system with realistic optical aberrations and wavefront control[9]. All of the systems in the downselect were evaluated over a $\Delta\lambda/\lambda_c = 10\%$ bandpass of 523 - 578 nm. The results were converted to science performance metrics[10] (e.g., number of planets that could be detected and characterized in a given time span). Each coronagraph had an advocate who provided the design (focal and/or pupil plane mask parameters, wavefront remapping functions) to be evaluated. The selected techniques had to be: 1) capable of operating within a 10% bandpass somewhere between $\lambda$ = 430 - 980 nm; 2) be able to detect at least 10 radial-velocity-detected gas giant planets of between 4 - 14 Earth radii in reflected light with contrasts $>10^{-9}$ assuming 1.6 mas RMS of pointing jitter and a factor of 10 reduction in the background speckle noise from post-processing; 3) detect circumstellar disks with $6 \times 10^{-9}$ contrasts (equivalent Airy disc diameter surface brightness) at $r = 0.25$ arcseconds; and 4) have a predicted NASA technological readiness level[11] of 5 by 2017 and 6 by 2019. This process occurred during the second half of 2013.

The HLC and shaped pupil coronagraphs were chosen as the primary designs and were combined into what was called the Occulting Mask Coronagraph (OMC); both share the same optical layout. The PIAACMC, which was considered less technically mature but with promising potential performance, was designated a backup with lower development priority. The other coronagraphs were deemed too immature, complex, or low in performance to satisfy the need to be technologically ready by the upcoming decision deadline for the inclusion of a coronagraph on AFTA and will not be discussed further here.

Note on conventions used: The contrast of a field is defined here as the planet-to-star flux ratio when the peak pixel of the planet's point spread function (PSF) is equal to the mean per-pixel brightness of the field. Angular separations are specified in terms of $\lambda/D$ radians (the units of radians are usually implicit). One $\lambda/D$ is equal to 0.047 and 0.069 arcseconds at $\lambda = 550$ nm and 800 nm, respectively. When discussing finite spectral bands, we will refer to a band's center wavelength as $\lambda_c$.

## 2   Optical system overview

The AFTA telescope is an on-axis Cassegrain with additional optics required for imaging at off-axis field positions. The beam from the telescope is picked off from an off-axis location by fold mirrors and into a pair of optics that collimate and correct for off-axis aberrations. A pupil image is formed at/near a fast steering mirror (FSM), and subsequently again at a 48×48 actuator deformable mirror (DM). Between these two is an optic on a piston stage to provide focus correction. There are different OMC (Fig. 2) and PIAACMC (Fig. 3) layouts from this point onwards. In the OMC there are two DMs for full-field wavefront control (one at a pupil, the other 1.0 m downstream), while the PIAACMC has only one for half-field control (an option chosen by its advocate to simplify the layout). In the OMC, after the DMs the pupil is reimaged onto a reflective shaped pupil mask located on a selection mechanism that allows a simple fold mirror to be used instead for the HLC. It is then focused onto a focal plane mask (FPM) after which another pupil image is created for the Lyot stop. A variety of shaped pupil and HLC FPMs and Lyot stops are provided using wheels. In the PIAACMC, the beam after the single DM is sent into the PIAA optics to be mildly apodized via compression and then to the FPM. A series of fixed Lyot stops at different locations along the optical axis are then applied. In both the OMC and PIAACMC layouts, after the Lyot stops there is a flip mirror that selects between the imaging and integral field spectrograph (IFS) channels, each with its own filter wheel. The layouts are complex - in the OMC layouter there are 31 reflections, including the telescope but excluding the IFS. Many of the optics are fold mirrors required to package the coronagraph in its limited allocated volume. Both channels utilize electron multiplied CCD (charge couple device) detectors. In the imaging channel a polarizing beamsplitter divides the beam into separate perpendicular polarizations that are imaged onto the same detector.

In the current configuration the imaging camera offers two 10% passbands centered at 465 nm and 565 nm to survey for planets and image circumstellar disks. It also has two 6% bands at 835 nm and 885 nm for in-and-out-of methane line imaging. The IFS will operate over three 18% passbands spanning 650 - 950 nm with a spectral resolution of R = 70. It will be used to spectrally characterize the planets identified earlier in the imaging channel (concurrent imaging and IFS measurements are not envisioned with this configuration as the wavefront control will be optimized for the bandpass of interest). To aid comparisons with the downselect results, all of the coronagraph results presented here are centered at 550 nm or 800 nm.

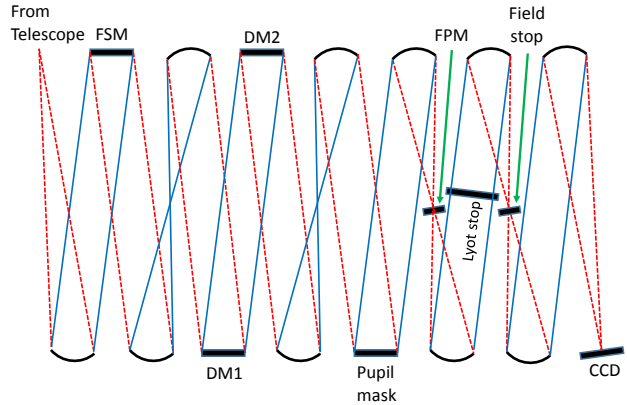

Fig. 2 - Conceptual layout of the post-downselect Occulting Mask Coronagraph composed of the shaped pupil and hybrid Lyot coronagraphs. Fold mirrors are not shown.

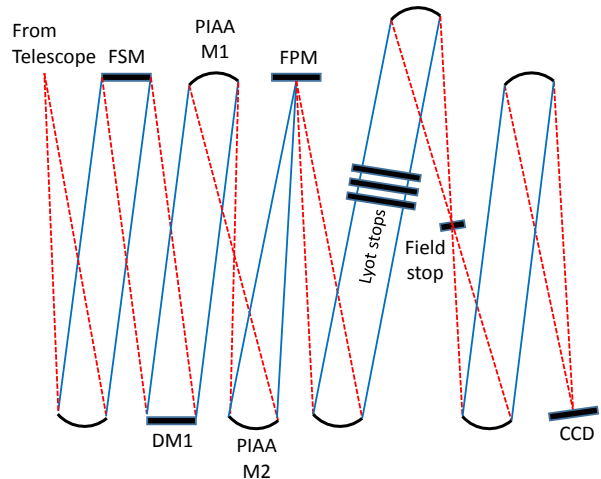

Fig. 3 - Conceptual layout of the post-downselect phase induced amplitude apodization complex mask coronagraph. Fold mirrors are not shown.

Each AFTA coronagraph will include a low order aberration sensing and control (LOWFS/C) system to maintain wavefront stability over time. In the HLC and SPC the FPM is reflective and directs the rejected stellar PSF core into a Zernike phase contrast sensor[12]. A patterned phase modulating coating on the FPM causes internal interference that transforms phase errors into measureable intensity variations when the reflected beam is imaged on a detector at a pupil. Because this does not work for the phase-only PIAACMC FPM, the starlight rejected by a reflective Lyot stop is reimaged onto a detector with some amount of defocus. Operating on the difference between successive measurements, the LOWFS determines changes in rapid variations (pointing errors) and longer term ones (focus, coma, astigmatism, etc.). The FSM corrects telescope body pointing errors down to a range of 0.4 - 1.6 mas RMS at kilohertz rates, while the focus correction mechanism controls focus variations at the rate of once every couple hundreds of seconds. The other low order aberrations are controlled using the deformable mirror that is at/near a pupil (the same DM that is used to create the dark hole) every few hundred seconds. The details of the sensing and control modelling procedure are discussed by Shi et al. in this volume.

# 3    Modelling procedure

A high performance coronagraph ideally suppresses the diffraction pattern of the star to a level below that imposed by the speckles of scattered light created by optical surface and coating errors. At the ~$10^{-9}$ planet contrast levels of interest these will dominate the background against which a planet is imaged. Wavefront control with DMs can be used to reduce the speckle intensity within a region around a star (the *dark hole*) that is limited in radius by the number of DM actuators spanning the pupil. Coronagraphs modify the behavior of these errors and thus can limit the control authority of the DMs, affecting how deep a dark hole the wavefront control system can provide. Pointing and vibration-induced wavefront jitters and polarization-induced aberrations also have an impact.

To accurately predict performance it is necessary to propagate a wavefront from surface to surface, picking up the phase errors (from figuring and polishing) and amplitude (coating non-uniformity) errors on each optic. Phase errors on non-pupil optics can create chromatically-dependent phase and amplitude errors in the image plane, an effect that must be captured in simulations as it presents a potential limit on the achievable broadband contrast[14]. Two DMs separated by an appropriate distance can be used to control both phase and amplitude errors to some degree over a 360° field, while a single DM can control them only over half that field (or a much smaller 360° field). A realistic model of the DM incorporating actuator influence functions is also needed. As will be shown, the DMs can also be used in conjunction with the pupil and/or FPMs to reduce the diffracted light, especially from obscured apertures[6].

The modelling process for each coronagraph began with the submission of a design by an advocate. Depending on the coronagraph (as detailed later) this may have included mask dimensions, FPM transmission pattern, DM surfaces settings, etc. The design was implemented in the modelling software and evaluated in an unaberrated system to verify that the performance matched that expected by the advocate. The field-dependence of the off-axis sources (planets) were determined. The system aberrations were then turned on and wavefront control used to reduce the resulting scattered light.

*3.1  Propagation and modelling software*

The use of a common model and propagation software allows the coronagraphs to be evaluated on as even a playing field as possible. AFTA optical modelling primarily utilizes the PROPER propagation software library[15]. PROPER is currently available for IDL (Interactive Data Language) with Matlab and Python versions in development (see *proper-library.sourceforge.net* for the latest version). PROPER includes near and far field diffraction algorithms with automatic propagator selection to handle surface-to-surface propagations. Phase and amplitude errors on each surface can be generated from power spectral density (PSD) or Zernike polynomial specifications. Complex obscuration patterns can be created with antialiased edges. It also includes a deformable mirror model utilizing measured actuator influence functions. In some cases where the PROPER routines are not sufficient, specifically PIAACMC, additional procedures have been developed to accurately represent unique parts of the coronagraph. PROPER has been used for a number of space and ground based coronagraph studies and development. Recently its accuracy was verified against more mathematically rigorous algorithms in a NASA Technology Development for Exoplanet Missions (TDEM) study[16-19], and it was validated in coronagraphic testbed experiments[20].

The AFTA coronagraph layouts, one for HLC+SPC and another for PIAACMC, are unfolded into a linear sequence of optics (PROPER does not do any ray tracing, so tilted or off-axis properties of optics are not represented explicitly). The propagation distances between, and effective focal lengths of, the optics are specified within a program (the *prescription*) composed of calls to PROPER routines that in total represent the system. Powered optics are represented as ideal thin lenses.

The wavefront sampling and computational grid sizes are chosen as a compromise to provide adequate sampling of the system obscurations and DM actuator influence functions in near-pupil regimes and of the field near focus while preventing aliasing artifacts from wrap-around caused by the Fourier transforms used in the propagations. As will be discussed later, the pixelated shaped pupil masks are generated for specific samplings (512 or 1000 samples across the pupil). Broadband images are created by generating a number of monochromatic images at wavelengths sampling the bandpass and adding them together. Nine wavelengths are typically used in the simulations presented here.

It should be noted that PROPER does not model all aspects of wavefront propagation. It is limited in accuracy by the assumptions in the algorithms it utilizes, specifically paraxial Fourier-based angular spectrum and Fresnel propagation. Like most diffraction propagation programs, it models a linear sequence of optics, and any effects caused by off-axis optics must be represented as wavefront errors derived from ray tracing software if they are deemed important. PROPER also does not capture the electromagnetic interaction of the field with conducting materials that would require elaborate vector propagation calculations. A limited set of such calculations were done for the Terrestrial Planet Finder Coronagraph project, notably for the shaped pupil masks that had very narrow, sharp openings[21]. Based on those experiences, aperture-related vector effects are not expected to be significant in the AFTA coronagraphs considered here, though the variation in polarization-dependent Fresnel reflection coefficient effects due to angle of incidence is included using ray tracing software outside of PROPER.

*3.2 Aberrations and jitter*

Each optic in the AFTA model system has synthetic phase and amplitude error maps created from PSDs representative of those of similar real optics. PROPER generates a map by transforming the PSD curve into a two-dimensional radial function, taking the square root (to convert from power to amplitude), and then adding random phases in a complex valued array. The Fourier transform of this produces the map (Fig. 4). By the nature of this process the resulting map is isotropic without any spatially correlated structures such as polishing groves or ion beam marks that may be present in real optics. The subsequent scattered light field generated by such a map is therefore also isotropic and without spatially-correlated patterns such as rings or grids of diffraction patterns. In the AFTA simulations the maps are filtered to spatial frequencies below 64 cycles across the beam diameter to reduce the chances of Fourier-transform wrap-around from high spatial frequency components and to allow accurate interpolation of the maps to a given wavefront sampling. The primary and secondary mirror phase errors were generated using the PSDs derived from the interferometric measurements of the actual AFTA optics. Because these measurements have been deemed proprietary, for the sake of access by the community the synthetic maps are used for the simulations. Simulations using the actual maps (not presented here) show similar performance as with the synthetic ones. Both the primary and secondary are expected to be refigured and recoated anyway. Other powered optics are given RMS wavefront errors of 2.5 nm and flats are 1.5 nm. These are similar levels as the Gemini Planet Imager optics[22]. The DM facesheets and the unconventional PIAACMC optics have 5 nm RMS wavefront errors. The amplitude error map levels are based on proprietary analyses of samples from a coating chamber that was optimized for coating uniformity and have total RMS variations of <0.5%. Phase-induced amplitude errors dominate over the coating nonuniformity errors, which are well below the requirements for this mission and match those required for Earth-finding missions[14]. Both the phase and amplitude PSDs follow an $f^{-2.5}$ power law ($f$ is the spatial frequency).

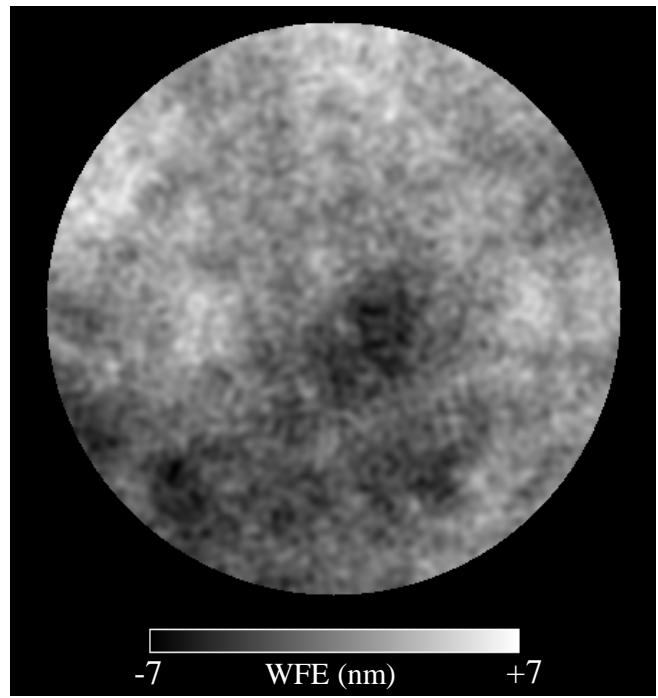

Fig. 4 - One of the synthetic phase error maps generated using a PSD specification that is used in the simulations to represent optical surface fabrication errors. Each optic has a different map.

To represent the effect of pointing jitter and a finite diameter star the source is offset by introducing a wavefront tilt at the primary mirror and then propagated through the system. This is done for a grid of 30 offset values at each wavelength, and the resulting intensity images are added together with weights defined by the convolution of a Gaussian jitter function and uniform stellar disk pattern. As will be shown later, jitter can have a significant impact on the achievable contrast floor, even after large telescope pointing errors are corrected by the FSM to expected levels of 0.4 - 1.6 mas (RMS per axis). At the current time only symmetrical jitter distributions have been considered.

Besides pointing (tip/tilt) errors, a coronagraph's response to other low order aberrations (e.g., focus, coma, astigmatism, etc.) is of particular concern. These are the aberrations most likely to change significantly over time due to thermal variations and vibrations of the optics. Any changes in them will result in changes in the scattered light pattern inside the dark hole and limit the effectiveness of post-processing techniques to differentiate planets from speckles. As a general indicator of a coronagraph's low-order aberration sensitivity, 100 picometers RMS of individual aberrations are separately injected into the system and propagated to create new images. The difference between the before-and-after-aberration maps is plotted versus radius to indicate which aberrations are of primary concern (these will be shown later). For a more detailed analysis of the sensitivities, individual low order aberrations are inserted into the system with a range of amplitudes. The resulting change in the electric field at the image plane (a complex value) is recorded for each. These fields are used in an error budgeting program[23] that allocates contrast performance contributions to various system properties (position stability of each optic, pointing offsets, etc.). The impact of a position error in one optic, for example, can be estimated by calculating via ray tracing the resulting low order aberrations, reading in the corresponding fields, adding them together and converting to contrast.

Because PROPER does not perform ray tracing itself, certain important system properties must be derived using software that does. These are characterized in a parametric manner (a function of wavelength or field position) so they can be applied in a PROPER model as a modification to the wavefront. The most important of these for the AFTA coronagraph is polarization[24], which is significant due to its fast primary mirror. Light incident on the optics with one polarization will reflect with a different low-order wavefront error than light with a perpendicular polarization. In on-axis systems like AFTA the primary polarization-induced aberration is astigmatism (with some tilt due to the off-axis location of the coronagraph). Without polarization filtering, the DM can be used to reduce low-order wavefront error down to the difference between the X and Y polarization aberrations. With filtering, one polarization channel can be fully optimized using the DM at the expense of contrast in the other polarization. Selected coatings can reduce but not eliminate the effect and not with equal effectiveness at all wavelengths. Using ray tracing software, the polarization-induced wavefront error maps in X and Y as seen at the pupil at the FSM were derived as a function of wavelength for the coatings assumed to be used on AFTA. The PROPER model reads in and applies these maps as appropriate. The coronagraphs have different sensitivities to polarization, largely due to differences in their behaviors with astigmatism. In the case of the shaped pupil in the IFS, where there is no polarization filtering, images are generated separately for X and Y polarizations and then averaged. Polarization effects were not included in any of the downselect analyses.

*3.3 Field (planet) PSF*

The sharpness of the coronagraphic PSF of an off-axis source (i.e., a planet) is a critical factor in computing the performance of a coronagraph in the presence of noise. Ideally most of the light would be in a sharp, narrow core, allowing for minimal background contamination and maximum effective throughput; deviations from this impact the signal-to-noise ratio of the system. The AFTA coronagraphs degrade the planet's PSF to some degree. The PIAACMC modifies the PSF the least, while the SPC alters the core the most. The tailored diffraction pattern of the shaped pupil redistributes a lot of starlight from the core out to large angles and broadens the core by about a factor of 2. The large-amplitude DM patterns integral to the HLC design push a significant fraction of light into the wings. Fig. 5 shows the field PSFs of the three coronagraphs. The encircled energy plots shown in Fig. 6 illustrate the severity of the loss of light from the core into the wings; in the HLC 80% encircled energy is not achieved until $r \approx 20\ \lambda/D$. Light from a faint source scattered to such large angles is essentially lost to the detector noise for practical exposure times. Reducing the amount of large angle scattering as a design optimization parameter is one means of improving the effective throughput of the system.

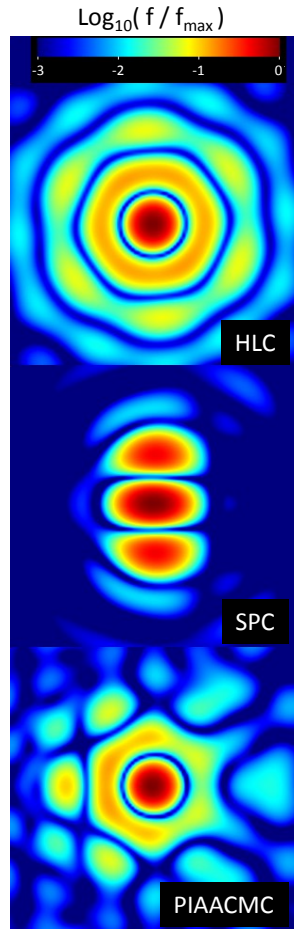

Fig. 5 - Field PSFs of the three coronagraphs displayed as peak normalized intensity over $\lambda$ = 523 - 578 nm. Each image is 10 $\lambda_c/D$ on a side.

A useful and intuitive metric for computing the signal-to-noise of a planet observation is the *PSF core throughput*, the fraction of a planet's light incident on the primary mirror that ends up in the core of the PSF (the region with values >50% of the PSF peak). This includes losses from pupil and focal plane masks but not from reflections, filter absorption, polarizers, detectors, etc. For AFTA without a coronagraph this is 34%. The *relative PSF core throughput* is the ratio of the PSF core throughput with the coronagraph to that without (34%) and is useful for comparing the transmission efficiencies of the different techniques to the ideal case. The *PSF core area* is also important for computing the contribution of the background to the noise measurement; this is 1650 mas$^2$ (square milliarcseconds) for AFTA without a coronagraph at 550 nm. The PIAA optics and large DM strokes used in HLC introduce some PSF field dependence, so these values are determined as a function of field position from the models by moving the star across the dark hole region in small steps by adding wavefront tilts at the primary mirror. The field radius from the star at which the PSF core throughput is 50% of its maximum is defined here as the *inner working angle* (IWA), which is a function of the FPM size (for SPC and HLC) or the effective amplitude modulation function (PIAACMC). The *outer working angle* (OWA) is set by the field stop (SPC), DM patterns (HLC), or designer-defined radius for wavefront control (PIAACMC).

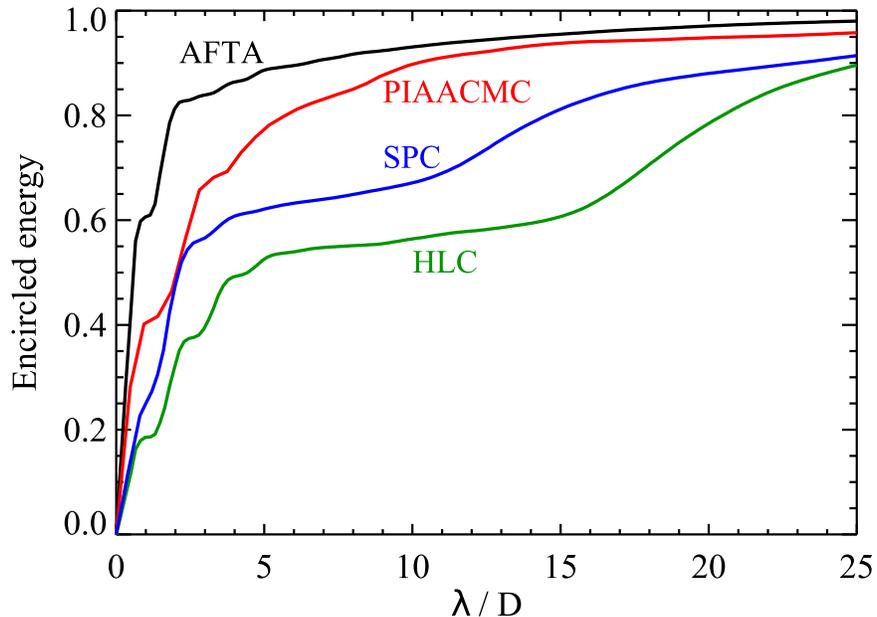

Fig. 6 - Encircled energy distribution versus radius of the field (planet) PSF of each coronagraph (revised designs) computed without the focal plane mask or field stop in place. The plot for the telescope without a coronagraph is also shown (AFTA). Each PSF is normalized by the total flux incident on the final image plane.

*3.4 Wavefront control*

Wavefront control with the DMs is used to reduce scattered light and generate a dark hole around the star. Prior to control the mean scattered light level in the hole has a mean contrast of about $10^{-4}$, which is dependent on the level of optical aberrations assumed in the system. The DM patterns that create the dark hole are computed using the Electric Field Conjugation (EFC) algorithm[25,26]. EFC has been utilized in the High Contrast Imaging Testbed (HCIT) at the Jet Propulsion Laboratory to generate dark holes with mean contrast levels below $10^{-9}$. EFC was used here because it has the best track record based on testbed results, its parameter space well explored, it is well documented, and does not require additional hardware that has to be modelled or added to the coronagraph in practice.

EFC assumes the coronagraph and DMs behave as a linear system. At its heart is the DM actuator response matrix that describes how the complex-valued electric field of each pixel within the dark hole changes for a given piston of each actuator. This matrix is generated by poking each actuator in the model system and recording the response in the computed field at each sensing wavelength.

The DM actuator settings that produce zero energy in the dark hole are derived using the pseudo-inverse of this matrix and the measured (or in the models, computed) dark hole electric field. To determine the chromatic behavior of the speckles and provide a broadband solution, the wavefront is measured at multiple wavelengths or sub-bands spanning the science bandpass (the DM response matrix contains the same wavelengths). Nine wavelengths were used for the simulations presented here. The DM response matrices can be very large; for instance, the matrix for the disk imaging shaped pupil coronagraph is over 13 gigabytes for dual polarization wavefront control (2 DMs with 1804 illuminated actuators each, 9 wavelengths, 2 polarizations, 12140 field pixels at 0.3 $\lambda_c$/D sampling, double-precision complex values).

Left unconstrained, EFC will attempt to solve an intrinsically nonlinear problem with a linear solution in one step using large DM strokes that actually degrade the contrast. To prevent this, a Tikhonov regularization of the DM response matrix is used[25,26] with identical constants applied to all actuators. Multiple regularization values are tested via trial-and-error to determine the optimal one for each case. With the best regularization the dark hole solution usually converges ($<10^{-11}$ contrast change per iteration) within 25 iterations. In reality more iterations would be required due to discrepancies between the model that generated the DM response matrix and the actual system, but in this all-simulation environment the two are the same unless purposefully altered.

The weight of each field point can be individually set during the EFC process to guide it to a better solution based on whatever a priori knowledge may be known about the behavior of the system. As demonstrated later, increasing the weights of field points at small radii from the star can improve the contrast not just there but over the full field. Low spatial frequency aberrations, which dominate at small angles, tend to impact the entire dark hole field due to their larger amplitudes relative to higher frequency errors and by the nature of the coronagraph. Without selective weighting EFC may find a solution for reducing light at large angles by introducing high spatial frequency patterns on the DMs when a better (in terms of contrast and/or bandwidth) solution is possible using low frequency corrections. Such weighting may even be introduced into the coronagraph design optimization process. We use here weighting functions when and as specified by the coronagraph designers.

In cases where the polarization is not filtered (e.g., the shaped pupil for the IFS), the fields are computed separately for X and Y polarizations (the same 9 wavelengths are used for each). Both are fed to EFC to obtain a common solution that is a compromise between the two polarizations.

In the simulations presented here for the aberrated AFTA system, the DM was used to take out ('flatten') the phase error of the wavefront incident on the critical component of each coronagraph: for the SPC at the shaped pupil mask, for the HLC at the FPM, and for the PIAACMC at the 1$^{st}$ PIAA mirror. This was done by propagating the wavefront at the central wavelength through the aberrated system up to the specified optic (and then to a crisp image of the pupil in the case of HLC) and extracting the phase map. The first DM's surface was then fit to this map, including actuator functions, and the opposite pattern was put on. This reduces the starting field contrast from ~$10^{-4}$ down to ~$10^{-6}$ and typically results in better final contrast using EFC (Fig. 7). In reality the phase map would be derived from pre-launch or in-flight phase retrieval measurements, such as was done to measure the mid-spatial-frequency errors in the Hubble Space Telescope[27].

In a real system, where only intensity can be measured, the dark hole field is derived by putting a set of known patterns on the DM and measuring the intensity changes in the image plane[28]; knowing how each DM pattern alters the field (via a model) allows the phase and amplitude of each pixel to be computed. For expedience in the simulations, the computed monochromatic fields at different wavelengths spanning the bandpass are usually used directly rather than derived via DM probing, providing perfect knowledge of the wavefront in the image plane. This is a best-case estimate of the contrast after wavefront control. In reality, the probed fields would be measured in multiple finite bandpass filters narrower than and within the wavelength range of the broad bandpass being controlled.

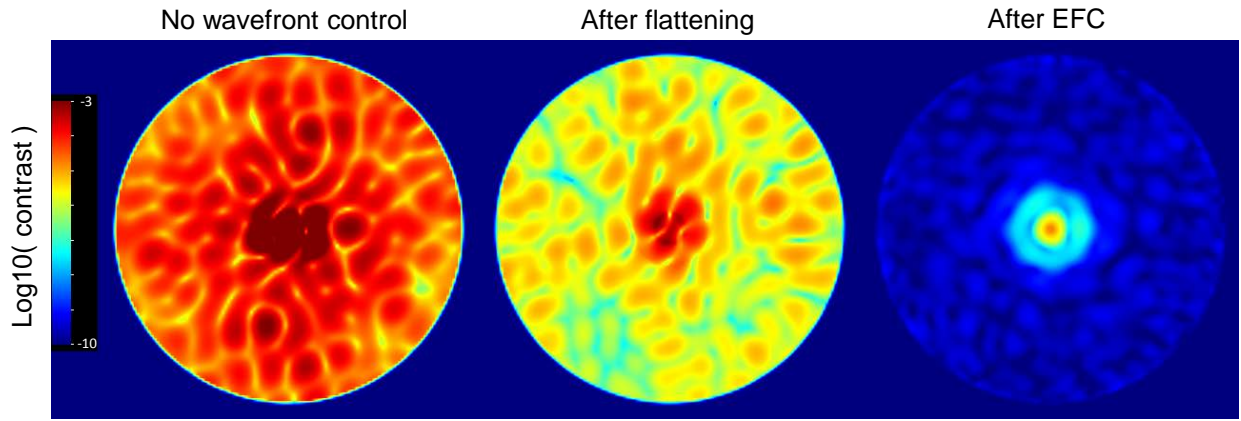

Fig. 7 - HLC dark hole contrast in the aberrated system (outer radius = 10 $\lambda_c/D$) over $\lambda$ = 523 - 578 nm (left) prior to any wavefront control; (middle) after setting the 1st DM to flatten the phase error incident on the FPM; (right) after phase flattening and running EFC.

It should be noted that the wavefront sensing implemented in these simulations neglects the effect of pointing and wavefront jitter; the system is assumed to be static while EFC is digging the dark hole. In reality all images, including those used for wavefront sensing, will be affected by jitter. However, because jitter is an incoherent phenomena (it is observed as the sum of multiple intensity images at different pointing offsets) and the wavefront sensing (via probing or otherwise) is sensitive to only coherent light, it is not seen as a wavefront error unless the form of the jitter changes between probes (e.g., the dominant jitter direction changes).

*3.5 Coronagraph science performance*

Specific characteristics of the coronagraph and the results of the simulations are used to derive the science performance metrics[29] (number of planets observed in a given time span). These include:

- The mean radial profiles of the broadband dark holes in terms of intensity and contrast relative to the star after wavefront control and application of different amounts of jitter. This is a component of the background noise.
- The PSF core throughput and core area as a function of angular separation from the star. The core region flux is essentially the only useful signal from the planet as the wings of the PSF will be well below the background noise in most cases.
- The FPM transmission pattern (multiplies the background signal from extended astronomical sources such as exozodiacal dust).

## 4 Shaped pupil coronagraph (SPC)

*4.1 SPC overview*

The SPC uses a binary pupil mask with tailored openings that control the pattern of diffracted light, distributing it outside of a desired dark hole region. The telescope obscurations are masked out as part of the design. The minimum IWA and best contrast that can be achieved are dependent on the size of the dark hole and the transmission loss that is considered acceptable. An advantage of a pure shaped pupil mask is that it is very insensitive to low order aberration changes, especially pointing errors (these simply cause an offset of the diffraction pattern, including the dark hole, but do not alter the contrast inside the hole). It is also achromatic, allowing it to be used over a broad bandpass, though the size of the diffraction pattern scales with wavelength. The downsides of the shaped pupil are its low transmission and broad field PSFs. Because of its large bandwidth it has been baselined for use with the integral field spectrograph in the current AFTA design studies. As will be described, the IWA and contrast of the SPC has been improved by adding a Lyot stop[30,31].

A variety of shaped pupils can be used to create dark fields in different areas. A "characterization" shaped pupil, intended for use with the IFS, provides the best IWA but at the cost of a limited OWA and azimuthal field coverage. Another mask intended for disk imaging creates a 360° dark field with a large OWA but also with a larger IWA.

The SPC designs presented here were developed at Princeton University by Neil Zimmerman, A. J. Riggs, Jeremy Kasdin, Robert Vanderbei, and Alexis Carlotti. See Zimmerman et al. in this volume for a more detailed description.

*4.2 Modelling the SPC*

The SPC pupil masks are provided by the designers as image files at the native resolution of the chosen mask pixelation. Previous shaped pupil designs used apertures with smooth edges and narrow, tapered openings that were both difficult to manufacture and to model with reasonable resolution. To avoid these issues the AFTA pupil masks are generated using finite width pixels that can be represented accurately in the computer and more easily manufactured. The mask is simply represented in the simulations as a binary array that multiplies the wavefront array. The wavefront sampling is chosen to match that of the mask, and the grid size is selected to prevent aliasing by Fourier transform wrap-around. In the downselect design a field stop is applied at an intermediate focus to prevent detector saturation and scattering by the backend optics. Because it simply blocks the bright light outside the dark hole it can also be represented as a binary mask that multiplies the wavefront.

In the post-downselect designs a Lyot stop is used, and this is also a binary mask. In the characterization SPC the field stop extends inwards to 2.5 $\lambda_c/D$, where the field is rapidly varying and some diffraction at the stop edges occurs. To more accurately capture the effect of the stop there the complex-valued field is interpolated to higher resolution and multiplied by a finely-sampled bow-tie-shaped mask. The result is then block averaged back down to normal sampling (real and imaginary parts separately). In the case of the disk imaging SPC field stop, which is an annulus and has a larger IWA, a simple anti-aliased representation at the default wavefront sampling is sufficient.

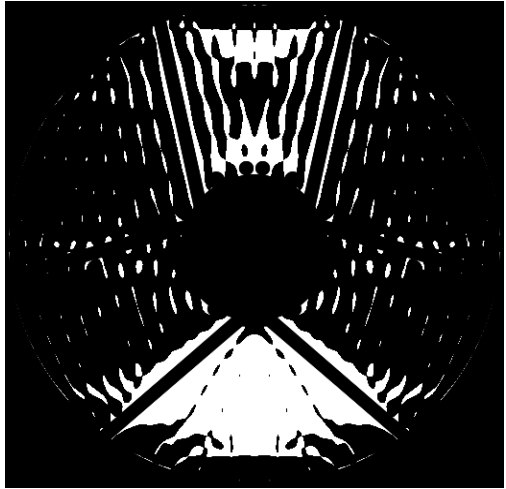

Fig. 8 - Shaped pupil mask used in the downselect.

### 4.3 SPC downselect design

The downselect characterization mask (Fig. 8) dark hole region extends from $r = 4 – 22.5\ \lambda_c/D$ in two 60° wedges on opposite sides of the star. A "bow tie" shaped field stop is used to prevent detector saturation. In an unaberrated system over a 10% bandpass it produces a mean contrast of $10^{-8}$ over the full field and $3 \times 10^{-8}$ over $r = 4 – 5\ \lambda_c/D$ (Fig. 9). The mask transmission is 22% relative to the obscured AFTA pupil, with a PSF core throughput of 2.7% (relative core throughput = 7.9%). At $\lambda = 550$ nm the core area is 2700 mas$^2$ (square milliarcseconds).

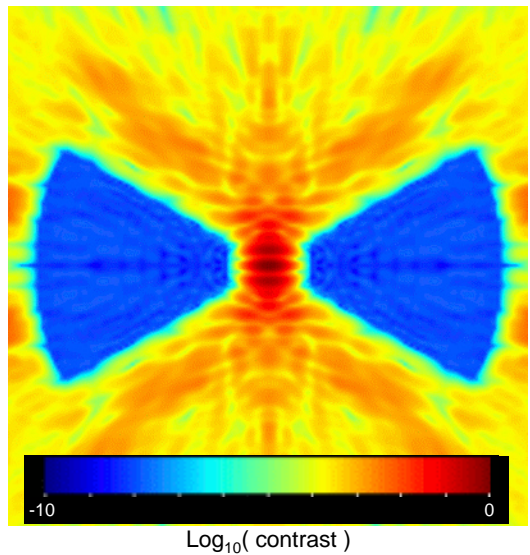

Fig. 9 - Downselect shaped pupil contrast in the unaberrated system without a field stop ($\lambda = 523 - 578$ nm, $\lambda_c = 550$ nm). The field is 50 $\lambda_c/D$ on a side.

Broadband EFC was run on the aberrated AFTA system over $\lambda$ = 523 - 578 nm ($\lambda_c$ = 550 nm), and the resulting mean contrast over the full dark hole was $7 \times 10^{-9}$ and from $r = 4 - 5\ \lambda_c/D$ it was $3 \times 10^{-8}$ (Fig. 10). The slight improvement in the aberrated system over the unaberrated one demonstrates the ability of wavefront control to partially contribute to diffraction suppression. When the 10% bandpass solution was used over a 20% bandpass ($\lambda$ = 495 - 605 nm) in the aberrated system, the mean field contrast degraded by only 30%, evidence of the pure shaped pupil's inherent achromaticity.

Using the EFC solution a broadband field was generated including 1.6 mas RMS per axis of jitter and a 1.0 mas diameter star. The contrast differences between the static and jittered fields were $<10^{-10}$, with higher changes in thin bands along the edges of the field stop caused simply by motion of the field. This reflects the jitter and low-order aberration tolerance of the pure shaped pupil design.

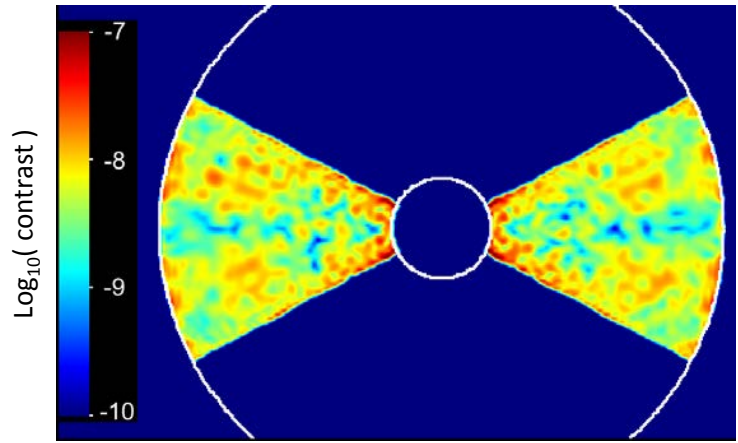

Fig. 10 - Downselect shaped pupil contrast in the aberrated system (no jitter) after EFC wavefront control ($\lambda$ = 523 - 578 nm, $\lambda_c$ = 550 nm). The circles are $r$ = 4 & 22.5 $\lambda_c/D$.

*4.4 Revised design with Lyot stop*

After the pure shaped pupil design was chosen in the downselect it was found that by adding a Lyot stop at a pupil after the field stop the performance could be improved, including having a smaller IWA. This configuration is somewhat analogous to the Apodized Pupil Lyot Coronagraph, but with a different apodization form. A large fraction of diffracted light that passes through the field stop gets concentrated inside the central obscuration and outside the aperture as viewed at the reimaged pupil. The Lyot stop there is simply an annulus blocking the inner and outer obscurations, with some oversizing. The shaped pupil is optimized for given field and Lyot stops. There are two current flight designs: the characterization mask (identified as SPC 20140902-1) provides an $r$ = 2.5 - 9 $\lambda_c/D$ field over two 65° opening angle sectors on opposing sides of the star over an 18% bandpass, while a disk imaging mask (SPC 20141007) provides a 360° field from $r$ = 6.5 - 20 $\lambda_c/D$ over a 10% bandpass and is intended for circumstellar disk imaging. In both cases the dark hole regions are defined by hard-edge field stops.

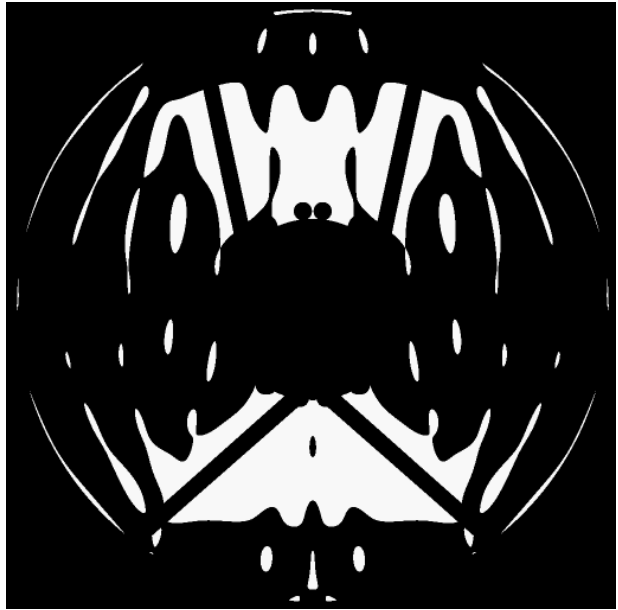

Fig. 11 - Revised shaped pupil mask.

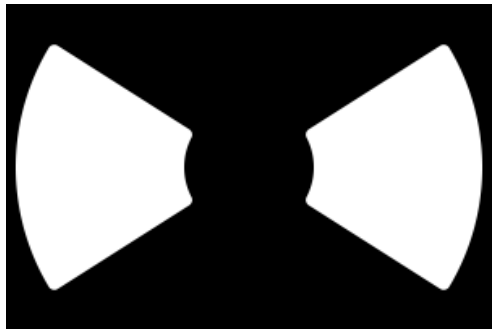

Fig. 12 - "Bow tie" field stop used as the focal plane mask for the revised shaped pupil in a 10% bandpass. The inner and outer radii are 2.5 and 9.0 $\lambda_c/D$, respectively.

*Characterization SPC mask*

As with the pure shaped pupil coronagraph, in the simulations the shaped pupil is applied as a binary mask that multiplies the wavefront (Fig. 11). Using a 4096 pixel diameter grid size and 1000 pixel diameter pupil mask the default sampling at the characterization mask's binary field stop is 0.24 $\lambda/D$ per pixel. This is interpolated to a higher sampling, the stop applied (Fig. 12), and the field rebinned. The field is then propagated to the next pupil image where the corresponding binary Lyot stop mask is applied.

The characterization SPC produces a field PSF with a FWHM area of 2600 mas$^2$ and a core throughput of 3.7% (relative core throughput is 11%). A considerable fraction of the light is diffracted to large angles by the mask. The PSF is tri-lobed (Fig. 5) with a central lobe FWHM of 2 $\lambda/D$ in one direction and 1 $\lambda/D$ in the perpendicular one. The IWA is $r = 2.8\ \lambda/D$.

The contrast results are given in Table 1. The mean contrast over an 18% bandpass ($\lambda$ = 728 - 872 nm, $\lambda_c$ = 800 nm) in an unaberrated system is $4 \times 10^{-9}$ over the full dark hole field and $1 \times 10^{-8}$ between $r$ = 2.5 - 3.5 $\lambda_c$/D. To see what additional gain in diffraction suppression that wavefront control could provide EFC was run on this SPC in the unaberrated system using the two DMs. At the suggestion of the Princeton team the region between 2.5 - 4.5 $\lambda_c$/D was given 5× greater weighting than the rest of the dark hole to force EFC to provide a solution better optimized for contrast at the IWA. This improved the mean contrast over the full dark hole to $2 \times 10^{-9}$ over the field and between $r$ = 2.5 - 3.5 $\lambda_c$/D.

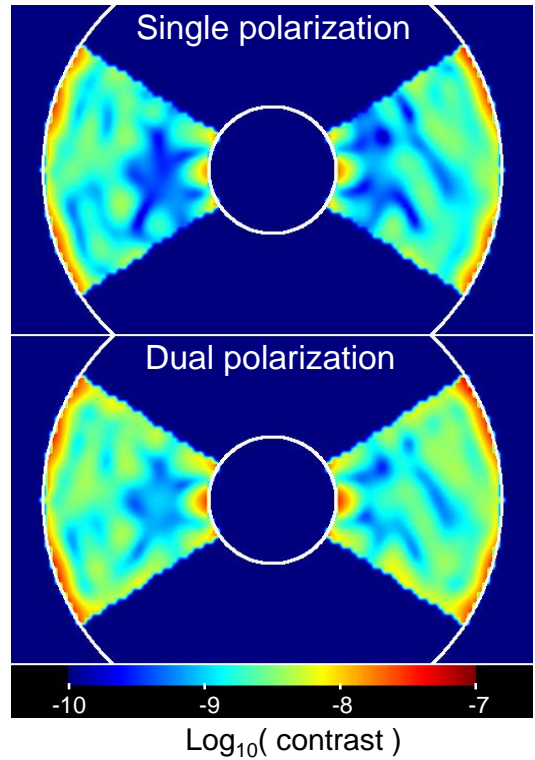

Fig. 13 - Contrast maps for the revised shaped pupil ($\lambda$ = 728 - 872 nm, $\lambda_c$ = 800 nm) after running EFC when (top) optimized for a single polarization, and (bottom) optimized for both. The circles indicate $r$ = 2.5 & 9.0 $\lambda_c$/D. No jitter has been added.

The unaberrated system EFC DM solutions were used as the starting point for another run of EFC in which the aberrations were included and the X and Y polarizations were optimized together (Fig. 13). Again, field-dependent weights were applied. EFC converged to a mean contrast of $4 \times 10^{-9}$ over the full dark hole field and $6 \times 10^{-9}$ between $r$ = 2.5 - 3.5 $\lambda_c$/D. As an experiment, EFC was run again optimizing and evaluating only the X polarization image, and the result was about 30% lower than the dual polarization values, close to the unaberrated contrasts. This shows that the contrast limit is primarily set by the shaped pupil and the polarization-induced aberrations rather than the optical fabrication errors.

The addition of jitter to the dual polarization solution shows a relative insensitivity to pointing. Compared to no jitter, the contrast degrades by about 50% with 1.6 mas RMS of jitter (Fig. 14). The aberration sensitivity curves shown in Fig. 15 reveal the SPC's tolerance of pointing (tip/tilt) and, important for polarization, astigmatism errors. It is most sensitive to coma and spherical aberration while being moderately sensitive to focus and trefoil. Its broad bandwidth and low sensitivity to pointing and polarization-induced aberrations make this SPC a good match to the long wavelength IFS (the polarization-induced aberrations increase at wavelengths longer or shorter than 550 nm).

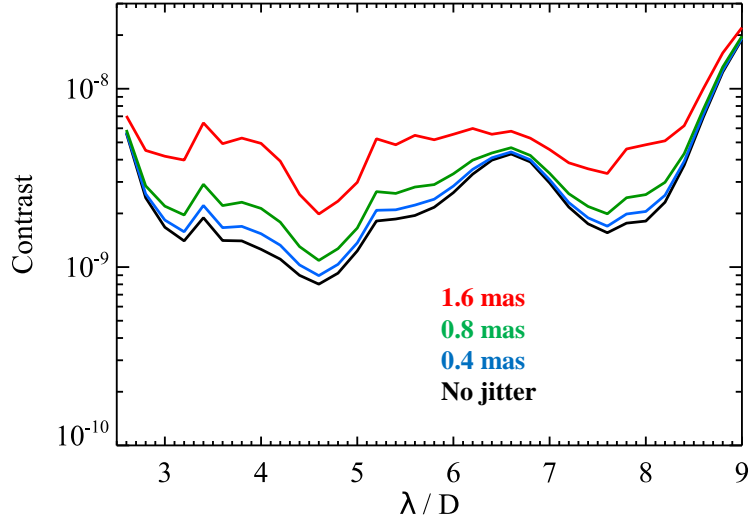

Fig. 14 - Revised shaped pupil radial mean contrast optimized for dual polarization (λ = 728 - 872 nm) with different amounts of RMS jitter (non-zero jitter results also include a 1.0 mas diameter star) added to the EFC solution.

Table 1. Revised SPC Results (λ = 728 - 872 nm, $\lambda_c$ = 800 nm)

| SPC Conditions | Mean Contrast | |
| --- | --- | --- |
| | 2.5 - 3.5 $\lambda_c$/D | 2.5 - 9 $\lambda_c$/D |
| Unaberrated, before EFC | $1 \times 10^{-8}$ | $4 \times 10^{-9}$ |
| Unaberrated, after EFC | $2 \times 10^{-9}$ | $2 \times 10^{-9}$ |
| Aberrated, single polarization, no jitter | $4 \times 10^{-9}$ | $3 \times 10^{-9}$ |
| Aberrated, both polarizations, no jitter | $6 \times 10^{-9}$ | $4 \times 10^{-9}$ |
| Aberrated, both polarizations, 0.4 mas jitter | $6 \times 10^{-9}$ | $4 \times 10^{-9}$ |
| Aberrated, both polarizations, 0.8 mas jitter | $7 \times 10^{-9}$ | $5 \times 10^{-9}$ |
| Aberrated, both polarizations, 1.6 mas jitter | $8 \times 10^{-9}$ | $6 \times 10^{-9}$ |

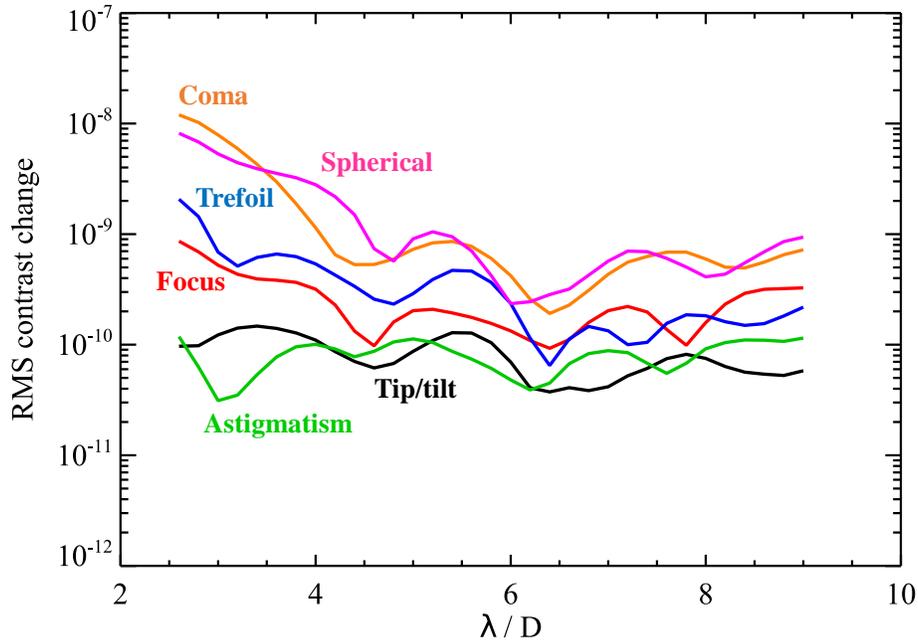

Fig. 15 - Contrast sensitivities of the revised characterization SPC calculated at λ = 550 nm for 100 pm RMS of wavefront change for selected aberrations. The RMS change computed over azimuth is plotted versus field radius. For non-circularly symmetric aberrations (e.g., coma), the aberration direction of maximum effect (e.g., X coma or 0° astigmatism) is plotted.

*Disk SPC mask*

The disk SPC mask is shown in Fig. 16. In the simulations the pupil mask is 512 pixels wide and a 2048 pixel diameter grid is used, producing the same 0.24 λ/D per pixel sampling at focus. The Lyot stop is a clear annulus between $r$ = 0.3 - 0.9 of the pupil radius. The maximum PSF core throughput is 7.1% (relative core throughput = 21%) and the maximum PSF core area is 1870 mas$^2$. The PSF FWHM is 1.0 $\lambda_c$/D, and the IWA is 6.8 $\lambda_c$/D.

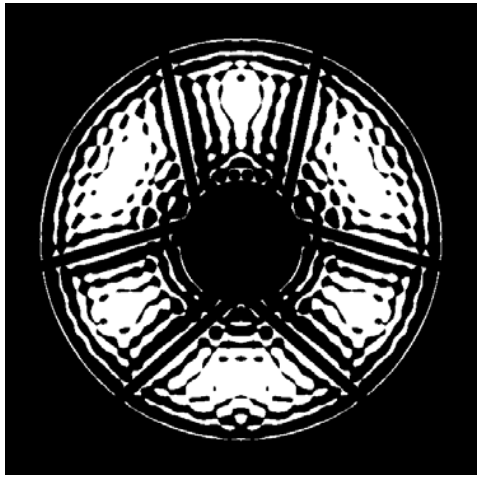

Fig. 16 - SPC disk imaging pupil mask.

Since the disk SPC is intended for the imaging channel it is evaluated from $\lambda$ = 523 - 578 nm ($\lambda_c$ = 550 nm). When running EFC field points between $r$ = 6.5 - 8.5 $\lambda_c/D$ are weighted 5× greater than the others. The contrast results are given in Table 2. Without aberrations this SPC produces a mean contrast field of ~4 × $10^{-9}$. As with the characterization SPC, EFC was first run on the unaberrated system, bringing the mean contrast down by a factor of ~3. With aberrations included, EFC was run separately optimizing for a single polarization channel and then for both together (Fig. 17). The resulting fields differ by ~$10^{-10}$ in mean contrast, so there is no advantage optimizing for just one polarization. This is advantageous since it allows imaging of disks in two polarization channels simultaneously. The simulations also show no significant contrast degradation (<5 × $10^{-11}$) with even 1.6 mas RMS of jitter.

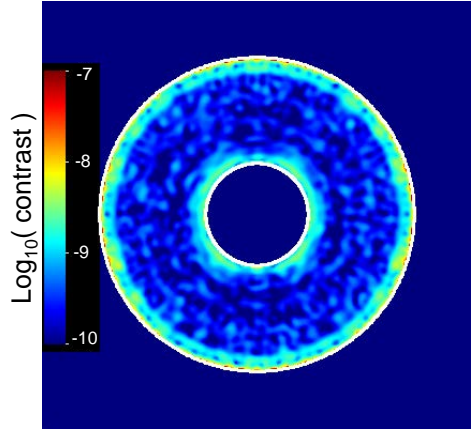

Fig. 17 - SPC disk mask contrast after running EFC optimized for both polarizations ($\lambda$ = 523 - 578 nm, $\lambda_c$ = 550 nm). The circles show $r$ = 6.5 & 20 $\lambda_c/D$.

Table 2. Disk Imaging SPC Results ($\lambda$ = 523 - 578 nm, $\lambda_c$ = 500 nm)

| SPC Conditions | Mean Contrast | |
| --- | --- | --- |
| | 6.5 - 7.5 $\lambda_c/D$ | 6.5 - 20 $\lambda_c/D$ |
| Unaberrated, before EFC | 5 × $10^{-9}$ | 4 × $10^{-9}$ |
| Unaberrated, after EFC | 2 × $10^{-9}$ | 1 × $10^{-9}$ |
| Aberrated, single polarization, no jitter | 2 × $10^{-9}$ | 2 × $10^{-9}$ |
| Aberrated, both polarizations, no jitter | 2 × $10^{-9}$ | 1 × $10^{-9}$ |
| Aberrated, both polarizations, 1.6 mas jitter | 2 × $10^{-9}$ | 1 × $10^{-9}$ |

## 5 Hybrid Lyot coronagraph (HLC)

### 5.1 HLC overview

The HLC[2,32] is a modification of the classical Lyot coronagraph that consists of an occulting mask located at an intermediate focal plane and a pupil mask at a subsequent pupil plane. In the HLC the FPM is a combination of a patterned amplitude modulator (a metal coating) with an overlaid phase modulator (a patterned dielectric coating). Both are simultaneously optimized to provide an acceptable compromise in IWA, contrast, bandwidth, low-order aberration tolerance, and throughput, with the wavelength-dependent characteristics of the materials included. The phase-and-amplitude modulating FPM provides better performance over broad bandpasses than previous amplitude-only designs.

Like most coronagraphic techniques the performance of the HLC is seriously impacted by obscurations in the telescope, especially the spiders. As part of the design optimization process the DMs are used to alter the wavefront to reduce the obscurations' diffractive effects, in a manner equivalent to the previously described ACAD technique[6,33] but using a different method of derivation. Pointing jitter is also included in the optimization. The resulting complicated pattern of actuator pistons has relatively large strokes (>200 nm), introducing a wave or more of peak-to-valley phase error. These patterns are an inherent part of the HLC diffraction suppression and would be used whether there were aberrations in the system or not. The large intentional wavefront errors, while improving contrast, do significantly degrade the field point spread function, scattering light from the core into the wings. Thus, the HLC has a fairly low effective throughput, more pronounced than would be expected simply from the loss of light at the Lyot stop. Beam walk on the second, non-pupil DM caused by pointing offsets also introduces a small field dependence in the PSF.

It should be noted that in the initial description[6] of the ACAD method it was asserted that the large wavefront errors from the high DM strokes invalidated the use of the Fresnel and angular spectrum propagation algorithms, which are used by PROPER. That conclusion was found to be incorrect, and it has been shown that those methods are actually accurate for the strokes considered here[34]. The DM patterns here were derived using these conventional propagators.

As in a classic Lyot coronagraph a Lyot stop mask is applied at the pupil image after the FPM to block light that was concentrated in and around the images of the obscurations. In the HLC the mask is also a diffractive element; some light is concentrated around the mask edges and ends up diffracting into the image plane where it destructively interferes to further improve contrast.

The HLC configurations described here were designed by Dwight Moody and John Trauger at JPL. See Trauger et al. in this volume for a more detailed description of the HLC.

*5.2 HLC modelling*

The HLC is relatively simple to model (though the optimization of its components is complex and computationally expensive). The HLC is specified by the designers as image files containing the complex-valued FPM modulation patterns for multiple wavelengths (including material thin-film effects), the Lyot stop pattern, and wavefront phase modulation patterns corresponding to the two DMs (including actuator influence functions). A 2048 × 2048 computational grid is used with a 336 pixel diameter pupil, which provides 7 samples across each DM actuator and an image plane sampling of 0.164 $\lambda$/D. The DM phase maps are added to the wavefront in the appropriate planes, the wavefront is multiplied by the FPM patterns at the intermediate focus, and is then multiplied at a subsequent pupil by the Lyot stop mask. The final fields were resampled to 0.3 $\lambda_c$/D.

*5.3 HLC downselect design*

The HLC design evaluated during the downselect (Fig. 18) had an IWA of 3.8 $\lambda_c$/D and an OWA of 13.2 $\lambda_c$/D over a 10% passband centered at $\lambda_c$ = 550 nm. The number of DM actuators across the pupil (N = 48) sets the maximum possible OWA (OWA$_{max}$ = N/2 = 24 $\lambda$/D), but optimizing over a smaller dark hole allows for improved contrast. The FPM and DM patterns were derived to provide reduced sensitivity to pointing errors assuming 1.6 mas RMS of jitter. The DM surface patterns had a peak-to-valley (P-V) range of 568 nm. An $r$ = 13.2 $\lambda_c$/D field stop at the intermediate focus after the Lyot stop was also used to block the bright light outside the dark hole.

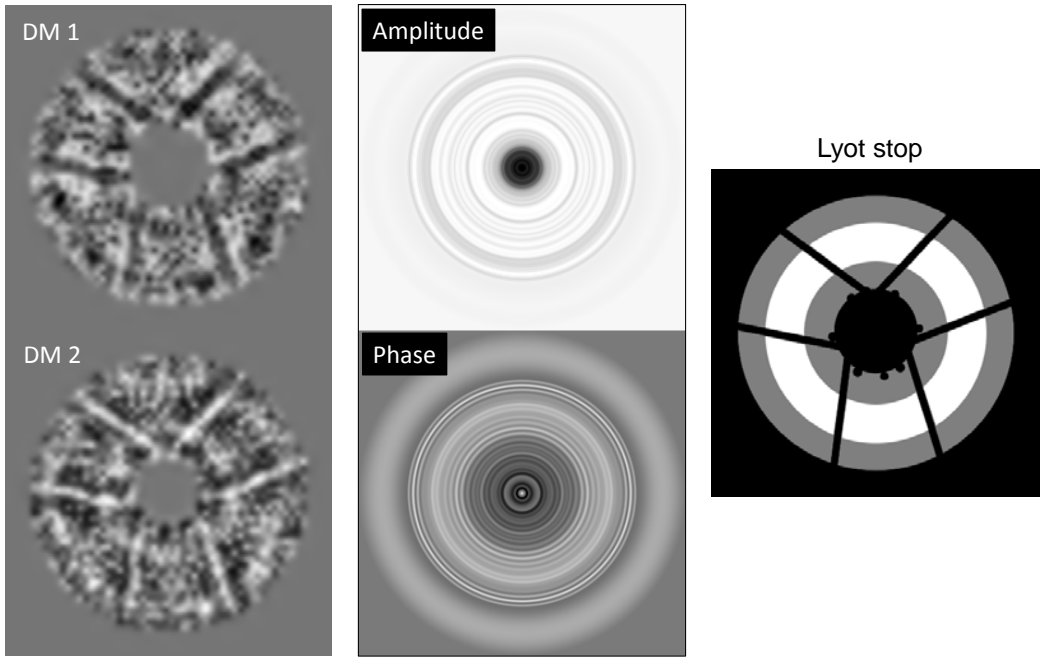

Fig. 18 - Downselect HLC components: (left) DM wavefront error patterns (1136 nm peak-to-valley); (middle) focal plane mask phase and amplitude modulation patterns at $\lambda$ = 550 nm (49 $\lambda$/D on a side); (right) Lyot stop pattern (grey) superposed on the AFTA obscuration pattern (black).

The large DM strokes introduced significant mid-spatial-frequency wavefront variations that, while beneficial to achieving good overall contrast when combined with the FPM and Lyot stop, resulted in a degraded planet PSF with only a small fraction of light in the core. The maximum PSF core throughput was 2.7% (relative core throughput = 7.9%) and the PSF core area was 1990 mas$^2$. This negates some of the advantages of having good contrast performance due to the low effective throughput, in a signal-to-noise sense (the core throughput, after accounting for losses from reflections, filters, polarization splitting, CCD efficiency, etc., would be <1%).

One effect of the optimization of the HLC for jitter tolerance was the deep contrast region near the IWA. This ~$10^{-10}$ contrast trough allowed the system to absorb some of degradation caused by jitter. When EFC was used in its default setup in the aberrated system, it tended to find a solution that provided fairly uniform contrast (~$10^{-9}$) over the full dark hole region, erasing the trough. When jitter was added to this solution, the contrast at the IWA was poor. To guide EFC to retain the trough the dark hole region from r = 3.6 – 6 λ/D was given 6× greater weighting than the outer radii. This was effective and allowed for better contrast after jitter.

The post-EFC HLC results are shown in Figs. 19 and 20 and Table 3. Without jitter the contrast at the IWA was $6 \times 10^{-10}$ but increased to $\sim 10^{-8}$ there and over most of the field with 1.6 mas RMS of jitter. Note that running EFC on the unaberrated HLC system would not improve contrast as it does with some of the other coronagraphs presented here because the DM patterns were already derived using an EFC-like optimization procedure.

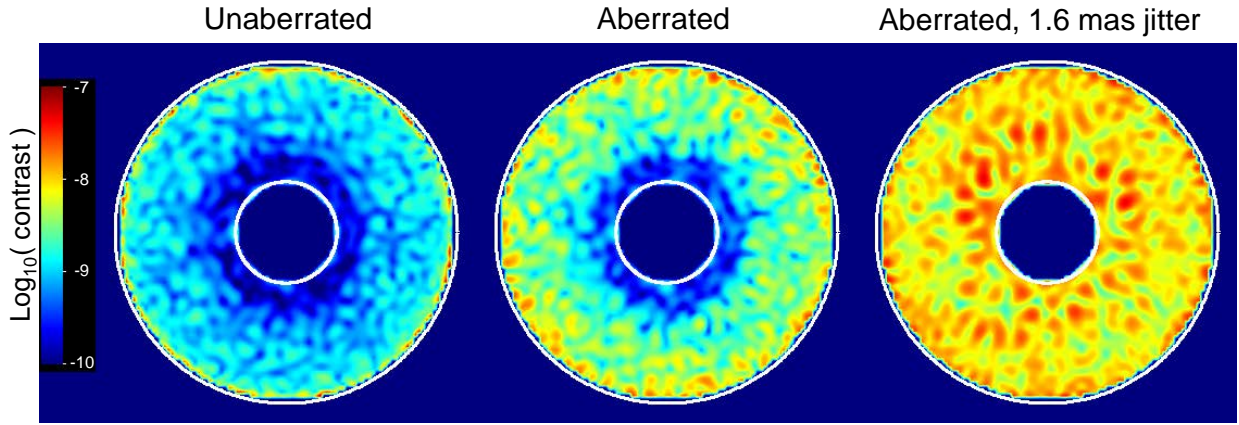

Fig. 19 - Downselect HLC contrast (λ = 523 - 578 nm; $λ_c$ = 550 nm); (left) unaberrated system (no wavefront control); (middle) aberrated system after running EFC; (right) with EFC solution and adding 1.6 mas RMS of jitter and 1.0 mas star. The circles represent $r$ = 4 - 13.2 $λ_c$/D.

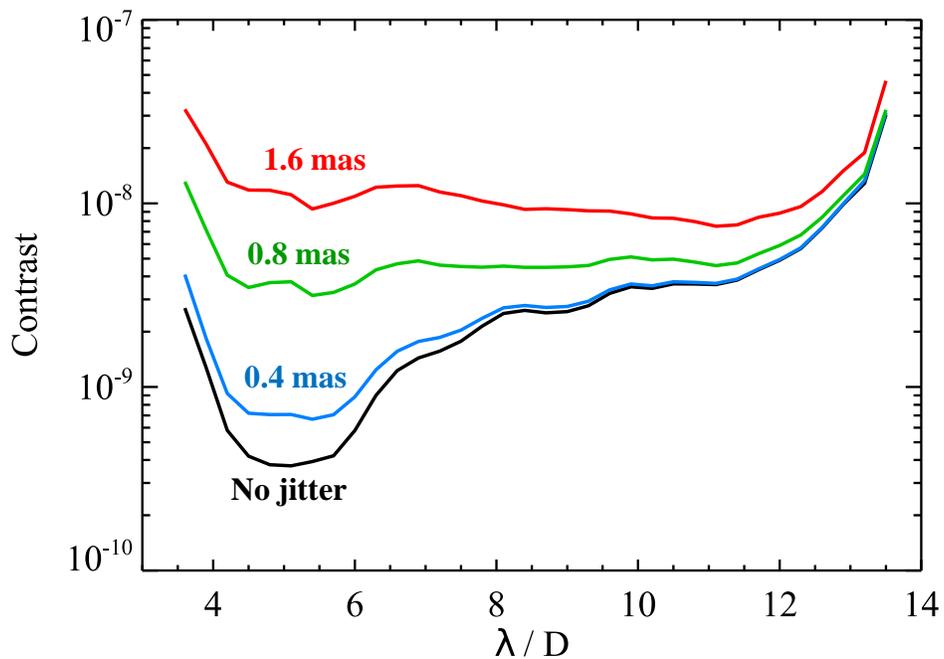

Fig. 20 - Downselect HLC radial mean contrast (λ = 523 - 578 nm) with different amounts of RMS jitter (non-zero jitter results also include a 1.0 mas diameter star) added to the EFC solution.

Table 3. Downselect HLC Results ($\lambda$ = 523 - 578 nm, $\lambda_c$ = 550 nm)

| HLC Conditions | Mean Contrast | |
|---|---|---|
| | 3.8 - 4.8 $\lambda_c$/D | 3.8 - 13.0 $\lambda_c$/D |
| Unaberrated | $5 \times 10^{-10}$ | $1 \times 10^{-9}$ |
| Aberrated, no jitter | $6 \times 10^{-10}$ | $3 \times 10^{-9}$ |
| Aberrated, 0.4 mas jitter | $10 \times 10^{-10}$ | $3 \times 10^{-9}$ |
| Aberrated, 0.8 mas jitter | $45 \times 10^{-10}$ | $5 \times 10^{-9}$ |
| Aberrated, 1.6 mas jitter | $141 \times 10^{-10}$ | $10 \times 10^{-9}$ |

*5.4 Revised HLC design*

Substantial optimization of the HLC design was made after the downselect with the goal of improving jitter tolerance, decreasing the IWA, reducing FPM complexity, and reducing DM stroke (Fig. 21). At the same time, the dielectric coating on the FPM was constrained so that the rejected PSF core light reflected by the metallic coating underneath it would have a suitable phase modulation for input into a Zernike phase contrast low-order wavefront sensor. The revised HLC design operates over a 523 - 578 nm bandpass ($\lambda_c$ = 550 nm) with an $r$ = 3.0 - 10.5 $\lambda_c$/D, 360° dark hole field defined by the DM patterns. The FPM now has a solid $r$ = 2.6 $\lambda_c$/D occulting spot with partial (~0.05%) intensity transmission, and the dielectric pattern does not extend beyond this radius.

The DM surface deformations have been reduced by about a factor of two from the downselect design to a P-V of 248 nm. With the lower DM strokes the PSF is sharper than the downselect design. The PSF core throughput is now 4.3% (relative core throughput = 13%), and the core area is 2100 mas$^2$. The IWA is 3.0 $\lambda_c$/D.

The HLC results are shown in Figs. 22-24 and Table 4. The unaberrated HLC provides a mean contrast of $2 \times 10^{-10}$ from $r$ = 3 - 4 $\lambda_c$/D and over the full field. With system aberrations EFC produces a mean contrast in a single polarization of $6 \times 10^{-10}$ from $r$ = 3 - 4 $\lambda_c$/D and $3 \times 10^{-10}$ from $r$ = 3 - 10 $\lambda_c$/D. When this solution is evaluated for the orthogonal polarization, the contrast is about 6× worse (the difference would be greater in passbands longer or shorter of 550 nm due to the increase in the polarization-induced wavefront error). Optimized for both polarizations the mean contrasts are about 2 - 3 times higher, so it may be practical to observe simultaneously in both polarization at once in the 523 - 578 nm bandpass without a huge signal-to-noise loss over a single polarization.

Note that the speckle patterns are different in the two polarization channels when EFC optimizes for both. On ground-based telescopes, with poorer contrast and where the instrumental polarization signature is not as dominant, the speckle patterns look the same in both polarizations, and simultaneous observations in both channels are used to distinguish polarized sky sources (primarily dust disks) from the background speckles. This will not be possible with any AFTA coronagraph.

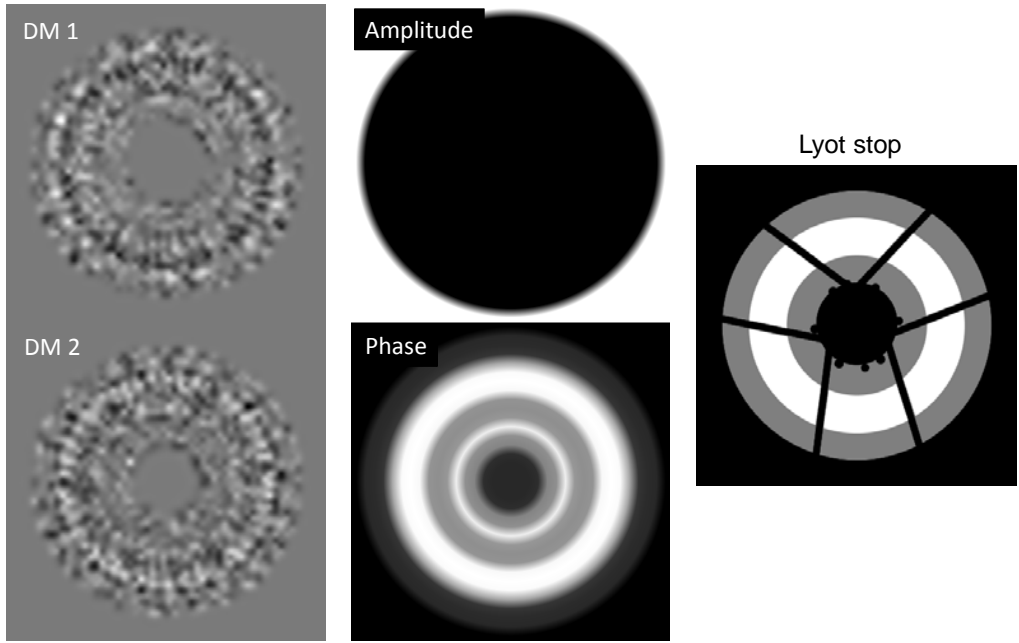

Fig. 21 - Revised HLC components: (left) DM wavefront error patterns (497 nm peak-to-valley); (middle) focal plane mask phase and amplitude modulation patterns at $\lambda = 550$ nm (6 $\lambda/D$ on a side); (right) Lyot stop pattern (grey) superposed on the AFTA obscuration pattern (black). The low spot in the FPM phase is intended for use by the Zernike phase contrast low order wavefront sensor.

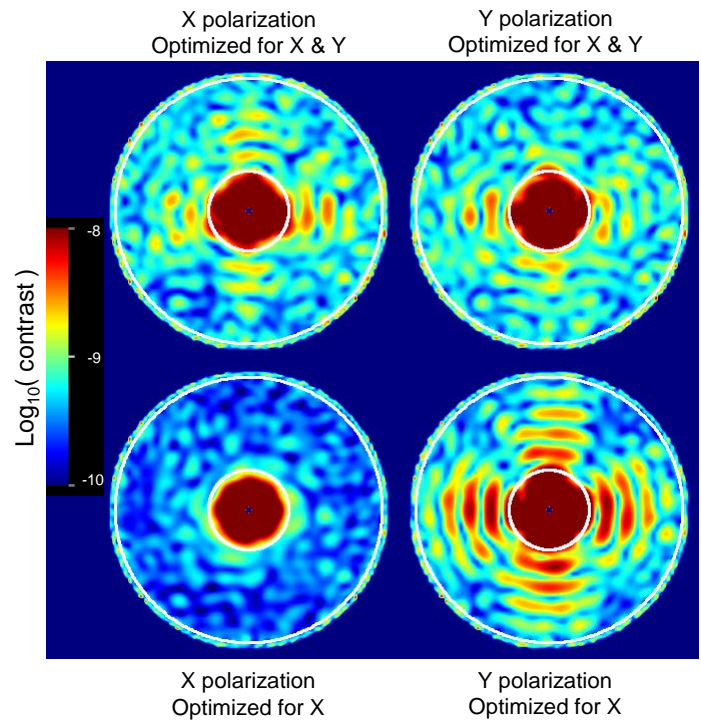

Fig. 22 - Revised HLC maps ($\lambda = 523 - 578$ nm; $\lambda_c = 550$ nm) for the aberrated system showing post-EFC contrast with 0.4 mas RMS jitter and 1.0 mas star in the X and Y polarization channels when EFC was optimized for (top) for both polarizations and (bottom) for only the X polarization. The cross-shaped patterns are due to the difference in astigmatism between the two polarization axes. Circles are $r = 3$ & $10$ $\lambda_c/D$.

When pointing jitter is added to the single polarization solution the contrast degradation is fairly low up to 0.8 mas RMS. With 1.6 mas RMS of jitter the contrast is about 7× worse at the IWA than the no-jitter case. However, at $r = 4\ \lambda_c/D$ its contrast is 7× better than the downselect design. Note that jitter introduces small-scale structures such as rings to the field. The aberration sensitivity plots (Fig. 25) show a nearly order of magnitude improvement in the tip/tilt and astigmatism tolerance at $3\ \lambda_c/D$ relative to the downselect HLC.

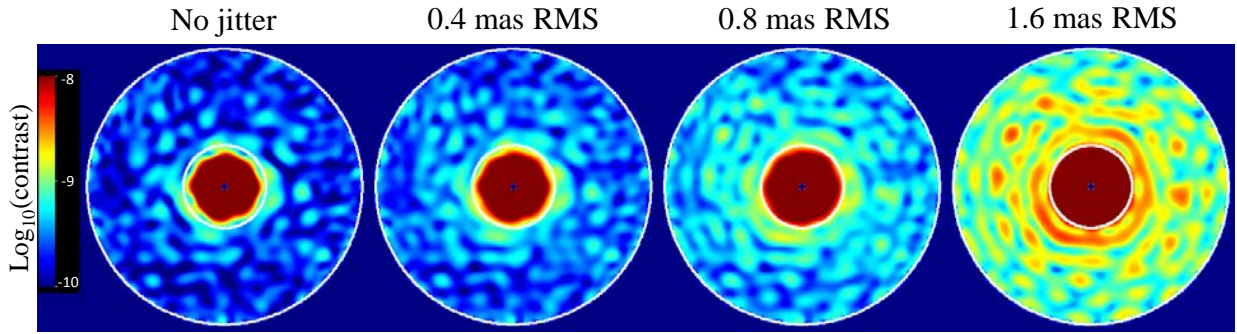

Fig. 23 - Revised HLC contrast maps ($\lambda$ = 523 - 578 nm; $\lambda_c$ = 550 nm) for the aberrated system showing post-EFC contrast in the X polarization channel (wavefront optimized for that channel) with 0.4 - 1.6 mas RMS jitter and a 1.0 mas star. Circles are $r$ = 3 & 10 $\lambda_c/D$.

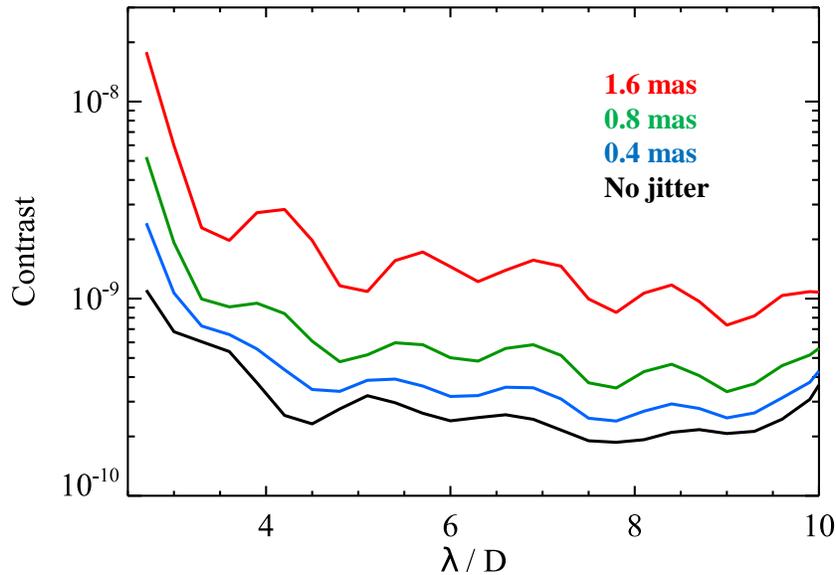

Fig. 24 - Revised HLC ($\lambda$ = 523 - 578 nm; $\lambda_c$ = 550 nm) mean azimuthal contrast versus field radius with different levels of jitter and (for jittered results) 1.0 mas star evaluated in and optimized for the X polarization channel.

An examination of the individual monochromatic fields that comprise the broadband dark hole simulation reveal no obvious correlations in the speckles with wavelength (Fig. 26). The speckle field essentially appears to "boil" when viewed over wavelength. In comparison, even the best ground-based coronagraphic systems, which operate at raw contrasts of >$10^{-6}$, are primarily limited by uncorrected phase errors that generate speckles with rather simple wavelength-dependent behavior (i.e., the speckle field appears to spatially grow with increasing wavelength). This allows for spectral differential imaging (SDI) post-processing techniques that rely on the predictable behaviors of the speckle pattern and planet (the speckle pattern grows with wavelength while the planet remains stationary).

Table 4. Revised HLC Results ($\lambda$ = 523 - 578 nm, $\lambda_c$ = 550 nm)

| HLC Conditions | Mean Contrast | |
| --- | --- | --- |
| | 3 - 4 $\lambda_c$/D | 3 - 10 $\lambda_c$/D |
| Unaberrated | $2 \times 10^{-10}$ | $2 \times 10^{-10}$ |
| Aberrated, single polarization, no jitter | $6 \times 10^{-10}$ | $3 \times 10^{-10}$ |
| Aberrated, single polarization, 0.4 mas jitter | $7 \times 10^{-10}$ | $3 \times 10^{-10}$ |
| Aberrated, single polarization, 0.8 mas jitter | $11 \times 10^{-10}$ | $5 \times 10^{-10}$ |
| Aberrated, single polarization, 1.6 mas jitter | $28 \times 10^{-10}$ | $14 \times 10^{-10}$ |
| Aberrated, both polarizations, no jitter | $10 \times 10^{-10}$ | $7 \times 10^{-10}$ |
| Aberrated, both polarizations, 0.4 mas jitter | $16 \times 10^{-10}$ | $8 \times 10^{-10}$ |
| Aberrated, both polarizations, 0.8 mas jitter | $20 \times 10^{-10}$ | $10 \times 10^{-10}$ |
| Aberrated, both polarizations, 1.6 mas jitter | $38 \times 10^{-10}$ | $18 \times 10^{-10}$ |

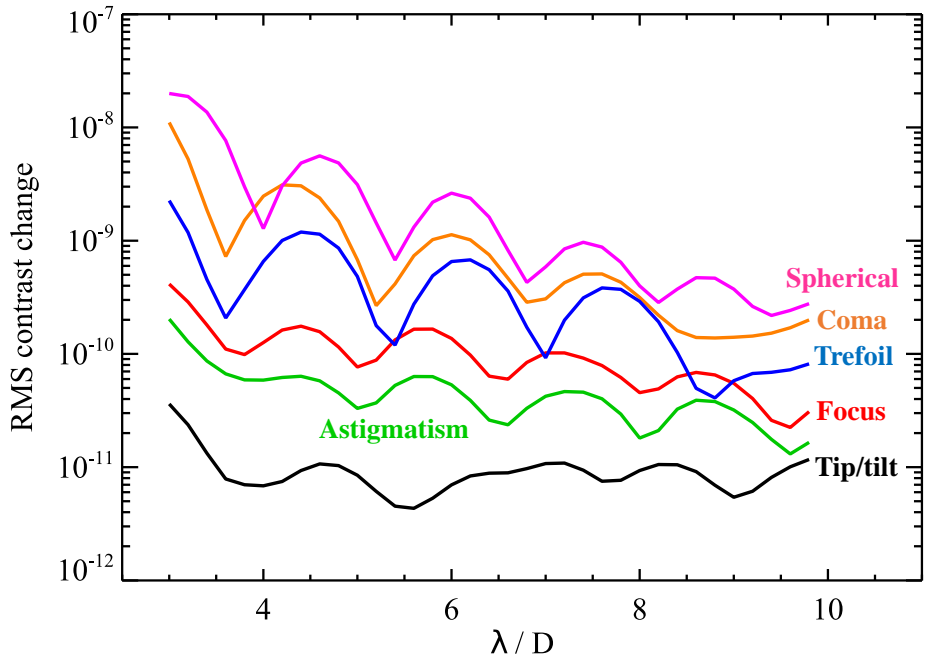

Fig. 25 - Contrast sensitivities of the revised HLC calculated at $\lambda$ = 550 nm for 100 pm RMS of wavefront change for selected aberrations. The RMS change computed over azimuth is plotted versus field radius.

In the case of a relatively stable space-based coronagraphic system like AFTA operating at ~$10^{-9}$ raw contrasts with sub-nanometer wavefront control the remaining speckles result from a complex mixture of phase and amplitude errors with different wavelength-dependent behaviors depending on how they were generated (e.g., phase-induced amplitude errors, DM residual print-through phase errors, etc.). The corresponding speckle fields therefore change rapidly over wavelength, hence the "boiling" (in the absence of wavefront control the phase-dominated speckle fields would appear to grow with wavelength). In addition, techniques like PIAACMC and HLC, which utilized strongly wavelength-dependent phase modifications at the focal plane, introduce further wavelength-dependent speckle variations. The application of SDI methods as they are use on ground-based telescope data on speckle fields like those shown here for AFTA will not work as the speckles have little wavelength correlation, as shown in Fig. 26.

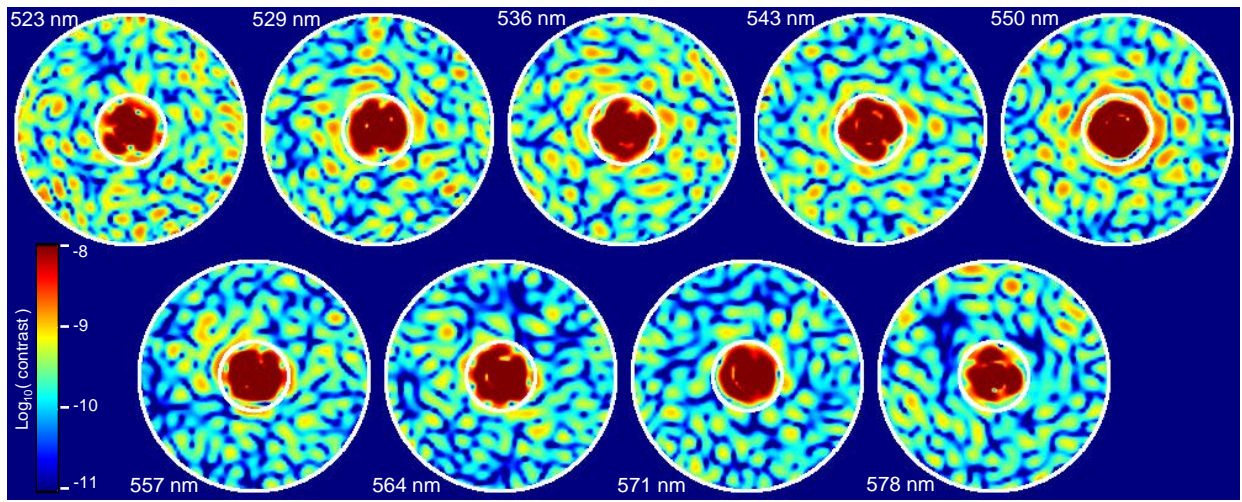

Fig. 26 - Revised HLC contrast versus wavelength evaluated and optimized for the X polarization channel (no jitter).

## 6 Phase induced amplitude apodization complex mask coronagraph (PIAACMC)

### 6.1 PIAACMC overview

The PIAACMC combines the nearly lossless beam apodization of a classical PIAA[35] coronagraph with a focal plane phase mask. A PIAA coronagraph relies on a pair of optics: the first (M1) compresses the beam into the desired pupil apodization profile and the second (M2) corrects for path length errors introduced by the remapping. After propagation to an intermediate focus the resulting PSF has a broadened core containing most of the light and significantly reduced wings. In the classical PIAA an opaque occulting spot is used to block the bulk of the starlight residing in the core, providing a high contrast field. The wavefront remapping results in a significant off-axis comatic distortion of the field PSF, so a reverse set of PIAA optics undistorts the wavefront, providing a sharp final PSF. The advantages of PIAA are a small IWA (a result of the magnification from the remapping) and high throughput (the apodization is done via beam reshaping rather than altering the transmission pattern with an absorbing mask). The classical

PIAA, however, is not tolerant of an obscuration pattern like AFTA's, resulting in unacceptably poor contrast.

The PIAACMC[3,36] variation replaces the opaque spot at focus with a small phase-modulating mask (the complex-mask, CM, part of PIAACMC) and uses much weaker PIAA optics. In the monochromatic case, the mask covers a portion of the stellar PSF core and shifts its phase relative to the rest of the PSF. As the beam propagates to a subsequent pupil interference causes starlight to diffract outside of the pupil as well as into the shadows of any obscurations where it is blocked by a Lyot stop. Because the bulk of the diffraction suppression is done by the phase mask, the apodization provided by the PIAA optics is much less aggressive, making the optical surfaces easier to fabricate. Whereas a classical PIAA system may provide magnifications in the focal plane of $3\times - 5\times$, the PIAACMC optics have magnifications more like $1.2\times$. As with a classical PIAA coronagraph reverse remapping optics can restore the wavefront distortion introduced during the apodization, but the less severe wavefront remapping of PIAACMC results in a fairly weak field dependence in the PSF, and they may be omitted for limited OWAs. In the absence of reverse optics the reimaged pupil after the FPM is not planar - different parts of the pupil are in focus at different locations along the optical axis. In this case there are multiple Lyot stops along the axis, each masking only those portions of the images of the obscurations that are in or near focus at each plane.

A simple phase mask consisting of a uniform-height spot of a suitable material (e.g., dielectric) works in monochromatic light, but in broadband the wavelength dependence of the phase shift, wavelength dispersion of the material, and the size dependence of the PSF necessitates a more complex solution. One method is to construct the phase mask with multiple concentric rings of material with different thicknesses optimized to provide the required contrast over a given bandpass. The joint optimization of the PIAA optics and these rings is complicated and numerically intensive, and it may include pointing error and other low order wavefront error tolerancing as constraints.

The PIAACMC designs presented here were developed by Olivier Guyon (National Astronomical Observatory of Japan/University of Arizona), Rus Belikov (NASA-Ames), and Brian Kern (JPL). See Guyon et al. in this volume for a more detailed description.

*6.2 Modelling PIAACMC*

When modelling a classical PIAA system the strong curvature of the optics results in a rapidly varying phase term preventing the use of conventional propagation algorithms like those used in PROPER. The wavefront cannot be finely sampled enough with practical array sizes to prevent aliasing of the phase. A new propagation method[37] was developed for PIAA that can handle the strong phase changes, but it is considerably slower than the Fourier-based techniques. An alternative but less accurate technique is to simply emulate the geometric remapping of the wavefront between the PIAA optics via interpolation, though this ignores any diffractive effects. Fortunately with the much weaker optics used in PIAACMC aliasing is not a concern and the usual Fourier methods work as long as the wavefront sampling is appropriately chosen.

The small, complex phase mask, however, requires some additional steps to be accurately represented. The wavefront is propagated at the default sampling to focus of the PIAA optics and is then Fourier transformed to a virtual pupil. Then it is propagated back to focus using a matrix Fourier transform[38] (MFT) at much higher resolution (approximately a thousand samples across the FPM). This field is computed only for the region within the phase mask. Two copies of this field are made. The FPM is applied as a map of phase offsets to one of these copies while the other is left unchanged. Both are then separately transformed back to the virtual pupil using an MFT. The unaltered field is subtracted from the original virtual pupil and the modified field is then added. The result is Fourier transformed back to focus and propagated as usual from there to the Lyot stops. The Lyot stops are simple binary masks that multiply the wavefront at the appropriate planes.

*6.3 PIAACMC downselect design*

The PIAACMC design submitted for the downselect (Fig. 27) included PIAA optics, a transmissive phase-modulating FPM, multiple Lyot stops, and reverse PIAA optics. Due to the spiders the PIAA apodization needed to be non-circularly-symmetric. This could be implemented in the PIAA optics with azimuthally-anamorphic surfaces and likely without too much fabrication difficulty. However, the computation of such surfaces is complicated and was beyond the time span of the downselect process, so instead a transmissive pupil mask with a weak, asymmetric amplitude modulation (~12%) was used in the simulations together with circularly-symmetric PIAA surfaces. The FPM consisted of 17 concentric rings extending to a radius of 2.5 $\lambda_c/D$. These were specified in terms of radii and heights of a given material (varying by a maximum of $\pm 1$ μm relative to the outer region). In the models the wavelength dispersion of the material was included. Four Lyot stops were applied after the FPM. This system was intended for operation with two DMs to provide a 360° dark hole around the star.

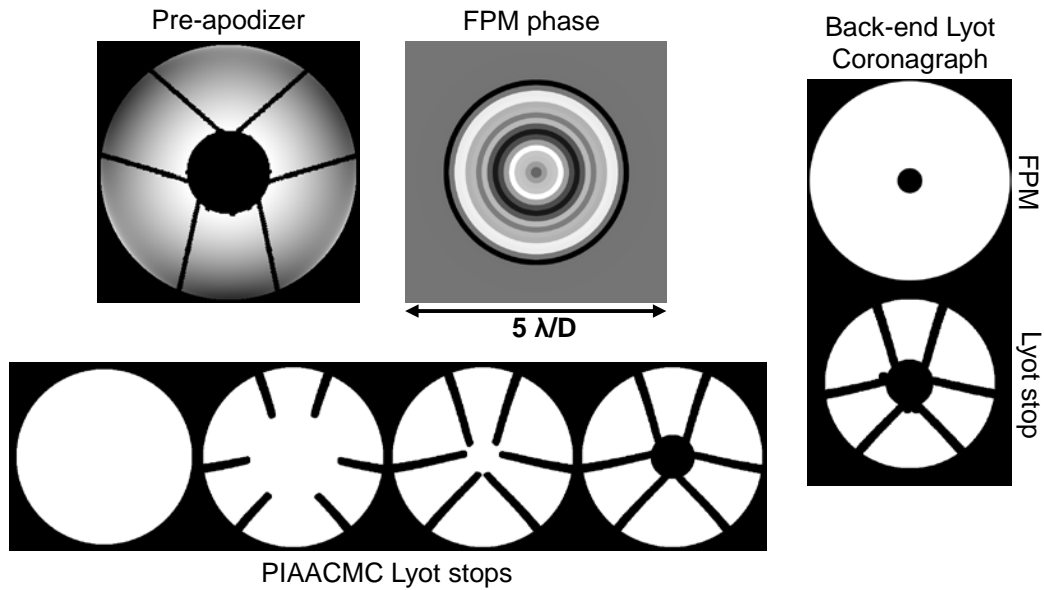

Fig. 27 - Downselect PIAACMC masks, including the back-end classical Lyot coronagraph. Notice that the compression of the beam by the PIAA optics shrink the apparent diameter of the central obscuration in the final PIAACMC Lyot stop.

Initial models showed unexpected residual starlight at small and large angles. It was not clear whether this was caused by the coronagraph or by numerical errors. It was rejected by adding to the back end a simple Lyot coronagraph with a ~1.7 $\lambda_c/D$ IWA and ~16 $\lambda_c/D$ OWA hard-edged focal plane occulter, followed by a Lyot stop that masked the obscurations. After this, the final stage was propagation through a reverse PIAA system to undistort the wavefront and then to the final image.

Because the PIAACMC Lyot stops do not significantly mask the outer edge of the pupil and a reverse PIAA system is used to undistort the wavefront, the transmission was high and the PSF shape was largely kept intact over the field. The peak PSF core throughput is 16% (47% relative core throughput), a high value compared to the other coronagraphs. The PSF core area is 1540 mas$^2$, slightly smaller than the non-coronagraphic PSF, an artifact of the multiple stages of optics reshaping the core. The IWA is 2.1 $\lambda_c/D$.

EFC wavefront control was used in the unaberrated system over an $r$ = 2.1 – 13.2 $\lambda_c/D$ field and 523 - 578 nm bandpass ($\lambda_c$ = 550 nm) and provided a factor of 2 - 3 improvement in mean contrast, down to $2 \times 10^{-8}$ from 2.1 – 3.1 $\lambda_c/D$ and $2 \times 10^{-9}$ over the full dark hole. EFC on the aberrated system increased the contrast level in the full hole to $2 \times 10^{-9}$ without changing the IWA contrast. The post-EFC results are shown in Figs. 28 and 29 and Table 5. The addition of jitter slowly increased the contrast level, with 1.6 mas RMS of jitter $1 \times 10^{-8}$ over the field and $9 \times 10^{-8}$ at the IWA.

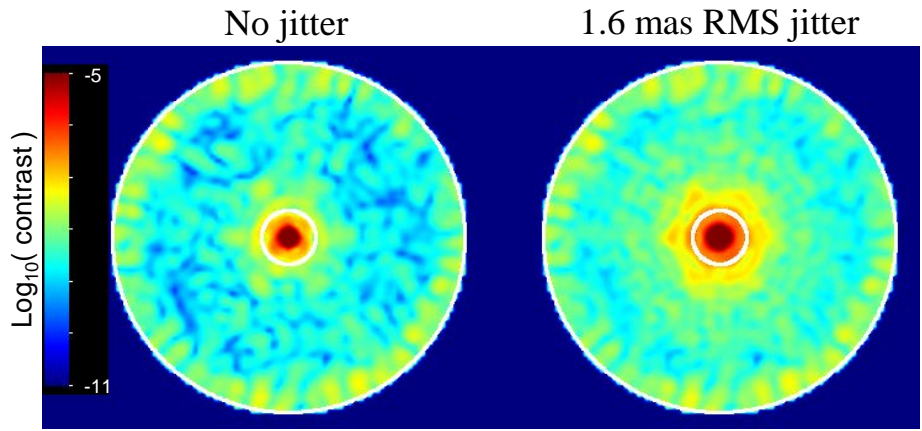

Fig. 28 - Downselect PIAACMC aberrated system post-EFC contrast maps ($\lambda$ = 523 - 578 nm; $\lambda_c$ = 550 nm). Jittered result also includes a 1.0 mas star. The circles are $r$ = 2 & 13 $\lambda_c/D$.

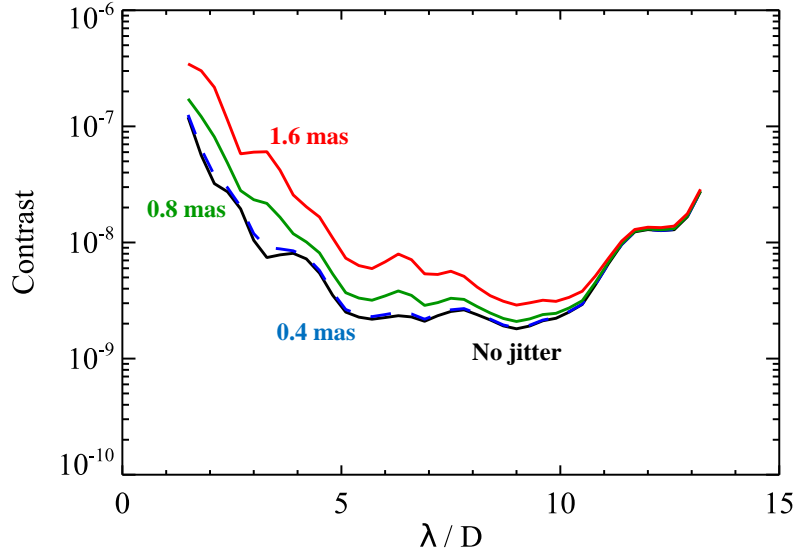

Fig. 29 - Downselect PIAACMC ($\lambda$ = 523 - 578 nm; $\lambda_c$ = 550 nm) mean azimuthal contrast versus field radius with different levels of jitter and (for jittered results) 1.0 mas star.

Table 5. Downselect PIAACMC Results ($\lambda$ = 523 - 578 nm, $\lambda_c$ = 550 nm, no polarization errors)

| | Mean Contrast | |
|---|---|---|
| PIAACMC Conditions | 2.1 - 3.1 $\lambda_c$/D | 2.1 - 13.2 $\lambda_c$/D |
| Unaberrated, before EFC | $4 \times 10^{-8}$ | $6 \times 10^{-9}$ |
| Unaberrated, after EFC | $2 \times 10^{-8}$ | $2 \times 10^{-9}$ |
| Aberrated, no jitter | $2 \times 10^{-8}$ | $6 \times 10^{-9}$ |
| Aberrated, 0.4 mas jitter | $2 \times 10^{-8}$ | $6 \times 10^{-9}$ |
| Aberrated, 0.8 mas jitter | $3 \times 10^{-8}$ | $8 \times 10^{-9}$ |
| Aberrated, 1.6 mas jitter | $9 \times 10^{-8}$ | $11 \times 10^{-9}$ |

### 6.4 Revised PIAACMC design

PIAACMC was designated as a backup technique to the baseline combination of the SPC and HLC. The high throughput and small IWA of PIAACMC were advantageous, but the complexity of the design, lack of a demonstrated hardware implementation, and poor performance in the presence of jitter made it too risky to baseline. After the downselect additional optimization was done to simplify it and improve the performance.

As with the downselect version, the current design (PIAACMC 20150322) includes weak PIAA optics, a focal plane phase mask, and multiple Lyot stops (Fig. 30), but it dispenses with the reverse PIAA optics and the backend Lyot coronagraph. An additional simplification is the use of a single DM for wavefront control, limiting the dark hole to one-half the area that would be possible with two DMs. The asymmetrical component of the PIAA apodization was removed, so the optics are now circularly symmetric (except for the off-axis component that is not included in the PROPER modelling and which may itself be eliminated using on-axis optics). The specified operating bandpass of this design is 523 - 578 nm ($\lambda_c$ = 550 nm). The phase mask is now a 3.2 $\lambda_c$/D diameter series of 22 concentric rings with azimuthal sectors of various heights (-339 to +355 nm) on a reflective surface. The addition of the azimuthal variations allows for more degrees of freedom for improving the bandwidth and jitter tolerance. The FPM is modelled in the same manner as the downselect version.

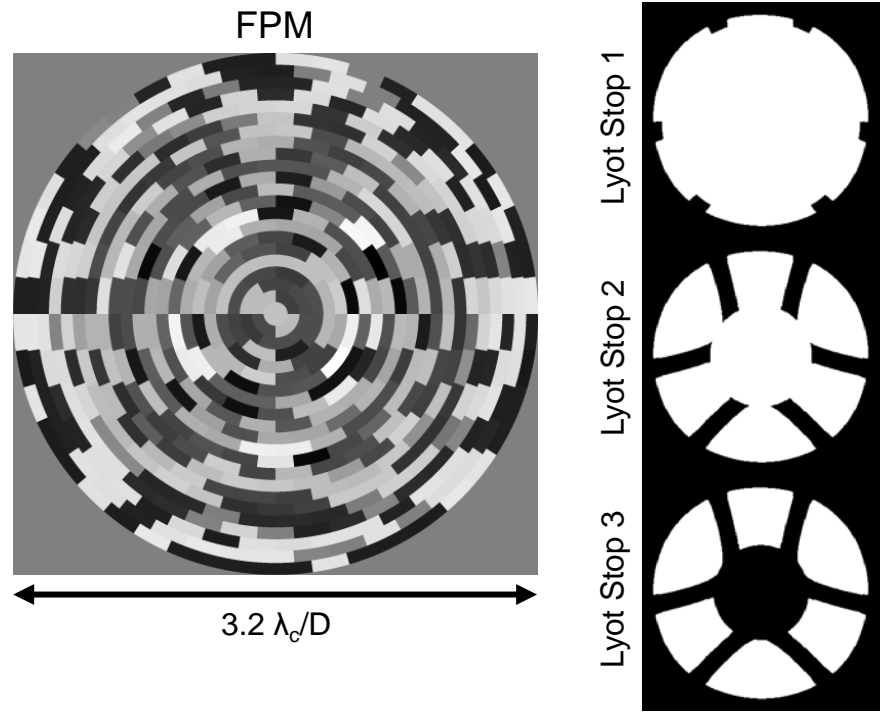

Fig. 30 - Revised PIAACMC focal plane mask surface height map and Lyot stops. The barrel distortion seen in the Lyot stop spider patterns is due to the remapping of the beam by the PIAA optics.

Due to the lack of a reverse PIAA, the field PSF varies by a small amount with the core throughput decreasing by about 20% towards the edge of the dark hole. The peak PSF core throughput is 14% (relative core throughput = 41%), the IWA is 1.3 $\lambda_c$/D, and the PSF core area is 1750 mas$^2$.

This design does not have a set OWA, but for the simulations presented here the dark hole was controlled over $r$ = 1.2 - 9 $\lambda_c$/D across one-half of the field around the star due to the single DM. In an unaberrated system prior to any wavefront control the mean contrast is $9 \times 10^{-9}$ from $r$ = 1.2 - 2.2 $\lambda_c$/D and $3 \times 10^{-9}$ over the full half dark hole. Running EFC on the unaberrated system reduces the mean contrasts by nearly a factor of 10. With aberrations included and optimizing for a single polarization there is little additional degradation in contrast (Figs. 31 and 32). With 0.4 mas RMS of jitter the contrast floor increased by factors of 5 - 9 near the IWA, and 1.6 mas RMS of jitter degrades contrast by nearly two orders of magnitude (Table 6). The sensitivity plots (Fig. 33) show that the tip/tilt tolerance is at least a few times worse than for the HLC but about the same as the characterization SPC. However, the SPC effective jitter tolerance is greater because it is sensitive only to jitter in directions along and near the axis of the dark holes (in the perpendicular directory the SPC field stop is essentially an infinitely large occulter).

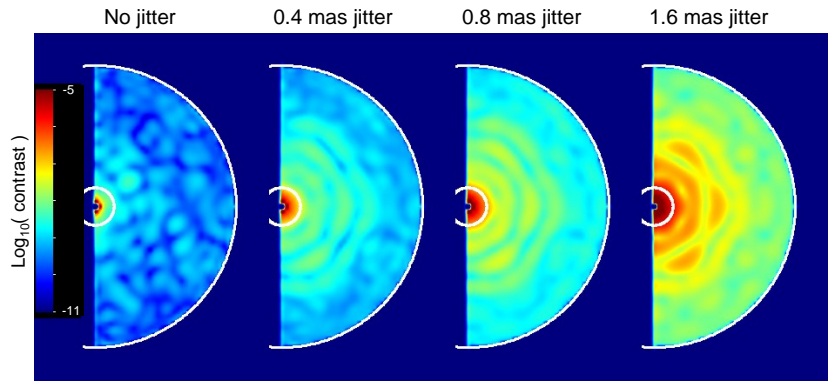

Fig. 31 - Revised PIAACMC contrast maps evaluated in and EFC optimized for a single polarization ($\lambda$ = 523 - 578 nm; $\lambda_c$ = 550 nm). Results with jitter also include a 1.0 mas star. The arcs are $r$ = 1.2 - 9.0 $\lambda_c$/D.

The PIAACMC astigmatism sensitivity is also about an order of magnitude greater than that for the HLC or SPC, and this reveals itself when the solution for one polarization axis is used to evaluate the contrast for the other axis. As shown in Fig. 34, the unoptimized axis has contrasts levels several orders of magnitude greater than for the optimized axis. This is primarily due to the astigmatism difference between the two polarizations. A compromise solution for both polarizations would obviously have unacceptable contrast, so PIAACMC must be used in a single polarization channel at all wavelengths.

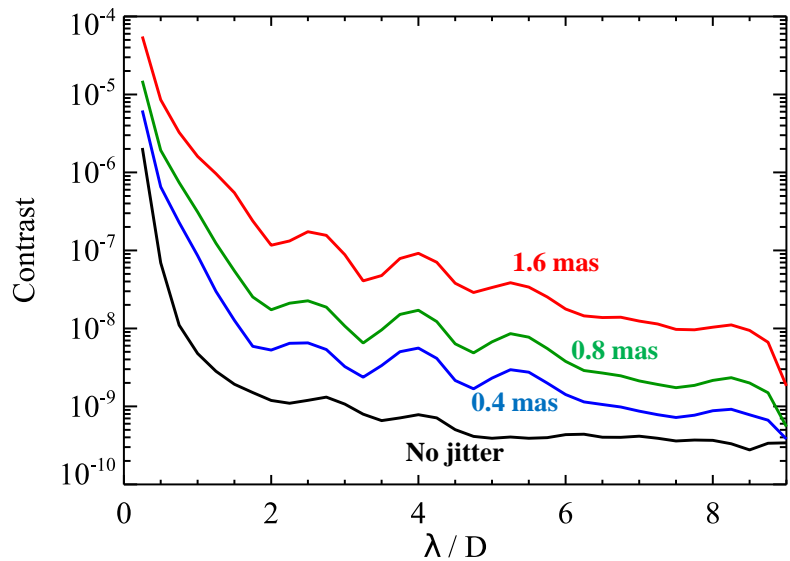

Fig. 32 - Revised PIAACMC ($\lambda$ = 523 - 578 nm; $\lambda_c$ = 550 nm) mean azimuthal contrast versus field radius with different levels of jitter and (for jittered results) 1.0 mas star evaluated in and optimized for the X polarization channel.

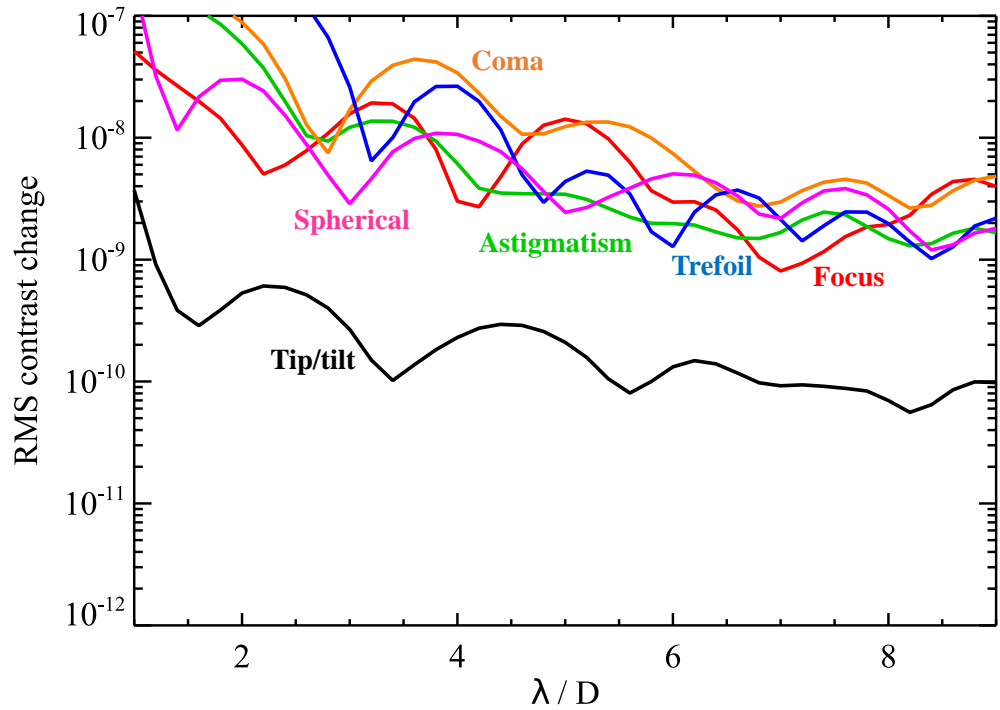

Fig. 33 - Contrast sensitivities of the revised PIAACMC calculated at $\lambda = 550$ nm for 100 pm RMS of wavefront change for selected aberrations. The RMS change computed over azimuth is plotted versus field radius.

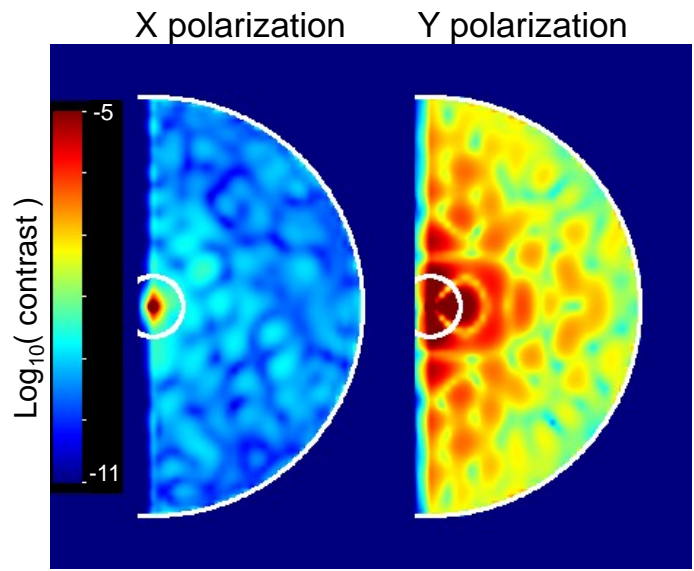

Fig. 34 - Revised PIAACMC contrast maps ($\lambda = 523 - 578$ nm; $\lambda_c = 550$ nm) optimized for the X polarization and evaluated in the X and Y polarization channel (no jitter). The arcs are $r = 1.2 - 9.0$ $\lambda_c/D$.

Table 6. Revised PIAACMC Results ($\lambda$ = 523 - 578 nm, $\lambda_c$ = 550 nm, single polarization)

| PIAACMC Conditions | Mean Contrast | | |
|---|---|---|---|
| | 1.3 - 2.3 $\lambda_c$/D | 3 - 4 $\lambda_c$/D | 1.3 - 9.0 $\lambda_c$/D |
| Unaberrated, no EFC | $9 \times 10^{-9}$ | $5 \times 10^{-9}$ | $3 \times 10^{-9}$ |
| Unaberrated, with EFC | $1 \times 10^{-9}$ | $0.6 \times 10^{-9}$ | $0.3 \times 10^{-9}$ |
| Aberrated, no jitter | $1 \times 10^{-9}$ | $0.8 \times 10^{-9}$ | $0.5 \times 10^{-9}$ |
| Aberrated, 0.4 mas jitter | $9 \times 10^{-9}$ | $4 \times 10^{-9}$ | $2 \times 10^{-9}$ |
| Aberrated, 0.8 mas jitter | $38 \times 10^{-9}$ | $12 \times 10^{-9}$ | $7 \times 10^{-9}$ |
| Aberrated, 1.6 mas jitter | $369 \times 10^{-9}$ | $70 \times 10^{-9}$ | $45 \times 10^{-9}$ |

## 7 Integrated modelling

Some of the most important questions about the performance of a coronagraph concern its stability during the long periods of time required for observations. The time to achieve a dark hole is on order of at least few hours on a bright star, and the field must be stable to within a fraction of the desired contrast, otherwise the wavefront control algorithm (e.g., EFC) will be always be chasing a fluctuating field and never converge. To distinguish the planet from background instrumental speckles the field must also be stable to within a fraction of the planet's contrast. The total exposure time needed to reach adequate signal-to-noise (e.g. *SNR* = 5) can be estimated based on some basic assumptions. The planet signal will be against a background that comes from many sources. The *SNR* is estimated as

$$SNR = \frac{S}{N} = \frac{r_{pl}\, t}{\sqrt{r_{noise}\, t + \sigma^2_{speckle}}}$$

where $S = r_{pl}t$ is the signal collected over integration time *t*, and *N* is the noise. The planet electron rate in the signal region is $r_{pl}$. It is a product of the stellar flux, the planet contrast, the collecting area, the instrument transmission, and the quantum efficiency. A simple definition of the signal region is the area within the PSF core. The random portion of the noise has variance that grows with time according to $r_{noise}\, t$ and includes shot noise from shot noise of the planet signal itself, shot noise of the (mean) speckle, shot noise of the incoherent (primarily zodical) background, detector dark current, read noise, and clock-induced charge. There is also a contribution to the variance that is not dependent on the integration time, written as $\sigma^2_{speckle}$: this is the spatial variance in the residual speckle structure, after post-processing, within the dark hole. For example in the case of 47 Ursae Majoris (47 UMa), a V = 5.0 mag G1V star with known RV planets at 3.3, 5.6 18.2 $\lambda$/D (at 500 nm), we estimate the middle planet's contrast to be $2.9 \times 10^{-9}$ for a typical observing phase and albedo of 40%. To image this planet, assuming the HLC coronagraph configuration with 0.8 mas residual pointing jitter and the baseline electron multiplication CCD (e2v CCD201 EMCCD) with sub-electron read noise, we arrive at an estimate of ~22,000 sec (0.26 day) to integrate up to a detection SNR of 5. In terms of integration times this is roughly the median case. Such durations are sufficiently long that thermal conditions can change significantly during them. Because post-processing of the images to reveal the planet signal is assumed to involve imaging another star for reference to calibrate the speckles, stability is needed not only during observations of the science target but also between and during the observations of the reference star.

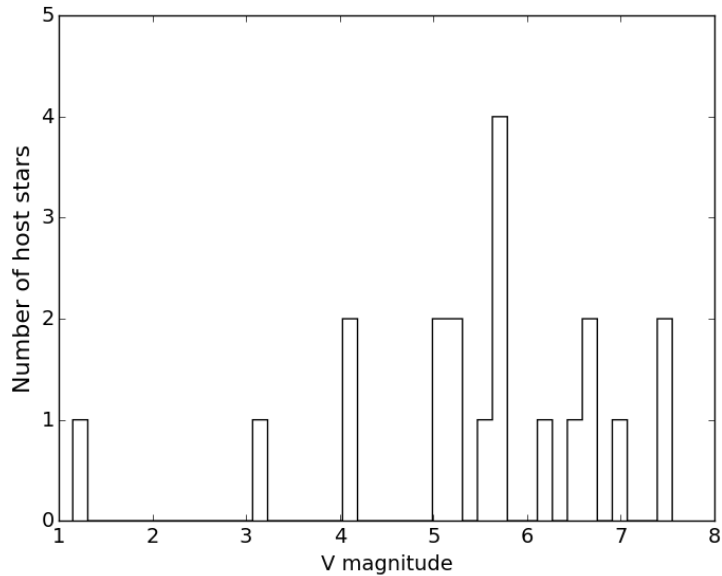

Fig. 35 - Distribution of V magnitudes for the candidate stars, selected as described in the text. There are 20 candidates total.

## 7.1 Observing Sequence Definition

In order to simulate the impact of thermal changes to various observing conditions, specific example observing sequences using possible actual targets are defined. While a complete target list for the mission has not yet been defined, a commonly used catalog is the known radial-velocity-detected planet hosts. For each of these we calculate the requisite integration time, per above assumptions, and sort the candidates according to integration time. Assuming an allocation of 30 days total integration time for the planet detection (imaging) portion of the mission (leaving the remaining allocation of one year of coronagraphic observations for spectroscopic characterization and disk imaging), we have 20 candidates for the HLC case. The histogram of planet host star brightnesses for this group is shown in Fig. 35. All of these 20 candidates are brighter than magnitude V = 8.

Target observation schedules will need to include constraints such as solar exclusion zones and solar incidence angles on the solar panels. If AFTA chooses an L2 orbit then the Earth exclusion zone will not be a strong constraint, while in a geosynchronous Earth orbit (GEO) it will.

The first devised observing scenarios assume a target star that is of average brightness and in the continuous viewing zone of the telescope. 47 UMa is the designated science target. At the brightness of this star the amount of exposure time needed to create a dark hole, including wavefront sensing via probing and evaluation of the field, would be prohibitive. It takes about 2000 sec just to get one photon per pixel in a $10^{-9}$ mean contrast field around a V = 5 star in a 10%

V band filter with the HLC. The strategy currently being considered is to select from the list of approximately 100 brightest stars in the sky the one nearest to the science target for generating the dark hole. This would mean that a typical bright star is roughly 20° away from any given science target. In our scenarios the bright star is $\beta$ UMa (Merak; A1IV, V = 2.4). Another star near the science target and with a similar color, 61 UMa (G8V, V = 5.3), is chosen to provide the reference images for post-processing subtraction of the speckles in the science frames.

The observing sequence is: 1) point at 61 UMa until the system reaches a thermal steady state (this is just for defining the starting state of the modelling sequence); 2) create the dark hole, or tune it up if it was created earlier, using the bright star $\beta$ UMa; 3) slew to the science target, 47 UMa, and take science observations; 4) slew to the reference target, 61 UMa, and take speckle field calibration images.

The moment the bright star is acquired in the coronagraph the LOWFS/C system is enabled to maintain the low-order wavefront stability using the FSM, focus correction mirror, and DMs. The dark hole is then iteratively generated, with each iteration consisting of probing using the DM to determine the wavefront error followed by a correction pattern being applied to the DMs. After the dark hole is achieved the telescope next slews to the science target and then the reference star. During these observations the higher order DM patterns are frozen at the settings derived with EFC (or similar algorithm) using the bright star. The LOWFS/C system, which is active during science and reference star observations, stabilizes the pointing and low order wavefront to the configuration state it had when the dark hole acquisition was completed on the bright star.

An important aspect of the timing specification is the *epoch* of the observation (Fig. 36). For our initial run (OS1), which was a first look at the speckle stability problem, the selected epoch provided the largest expected change in solar flux aside from Earth eclipse effects (which were specifically avoided). Furthermore, those times when the solar pitch angle exceeded 36º were excluded, consistent with solar array power requirements.

The timing of the observations requires estimation of the SNR achievable over an allocated amount of integration time. In the first pass of the study simple arguments based on estimates of SNR and experience in testbeds were used to set the exposure times for each of the three stars. Estimates of slew and settle times were based on engineering judgement from prior space instruments. The timing specifications for the observations in OS1 were made approximately suitable for imaging observations in one or two 10% bandwidth filters (but not long enough to take IFS measurements).

A more recent observing scenario, OS3, was developed that is simply OS1 with a different starting epoch to ensure that the slews do not occur at the same time as an Earth illumination transition. This is intended to better separate the effects of those disturbances. OS3 also utilizes an updated model of the system and was run with and without the Earth (as a stand-in for a potential L2 orbit). The OS3 timeline is detailed in Table 7.

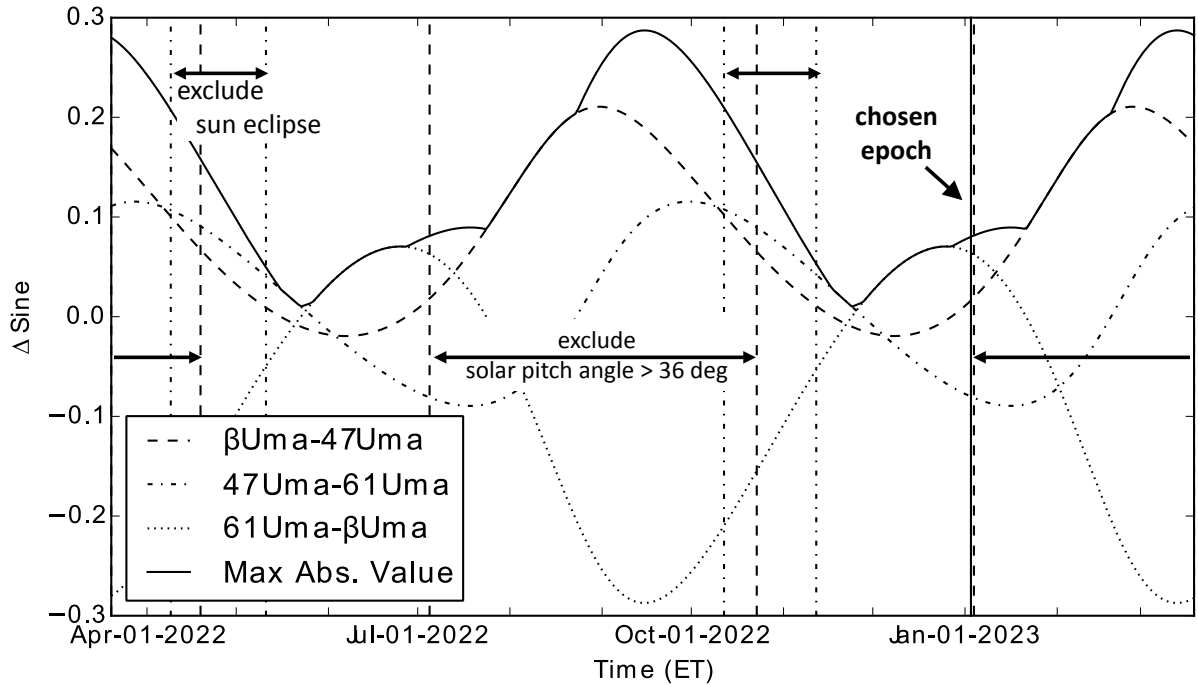

Fig. 36 - Choosing the observation epoch for the first OS1 modeling case. The curves are the differences of the sines of the angles to the various stars relative to the sun vector. The chosen epoch is within the 24 hour continuous viewing time but has the maximum thermal change from sun.

**Table 7. Timing definition for observing sequence OS3**

| Star | Function | From (hrs) | To (hrs) | Duration (s) |
|---|---|---|---|---|
| 61 UMa | Prior | 0 | 2.8 | 10,000 |
| $\beta$ UMa | Dark Hole | 2.8 | 9.7 | 25,000 |
| 47 UMa | Target | 9.7 | 33.3 | 85,000 |
| 61 UMa | Reference | 33.3 | 56.9 | 85,000 |

*7.2 Observing scenario thermal modelling*

During the scenarios the thermal loads on the instrument and spacecraft change depending on the angles to the Sun and Earth. The thermal finite-element model (FEM) computes at a specified cadence (e.g. every 2000-5000 seconds) the temperatures at all of the nodes defined for the spacecraft and instrument. These results are fed to a structural FEM which includes the thermal expansion coefficients between all of the nodes. Certain nodes in the FEM correspond to locations on the primary, secondary, and subsequent optics.

To analyze the effects of thermal/structural distortions on the optical performance of the telescope and coronagraph, the optical surface rigid body motions (RBMs) and surface deformations are determined from the structural FEM. The surface distortions are characterized by a set of standard

Zernike coefficients. The surface displacements and Zernikes modify an optical prescription of the telescope+coronagraph system that is then ray-traced to produce the wavefront aberrations seen at the FSM. These are then fitted by Zernike polynomials and are passed on to PROPER. The official AFTA thermal model results are provided by the Goddard Space Flight Center, while additional cases (e.g., rolling the telescope) are explored by the Jet Propulsion Laboratory.

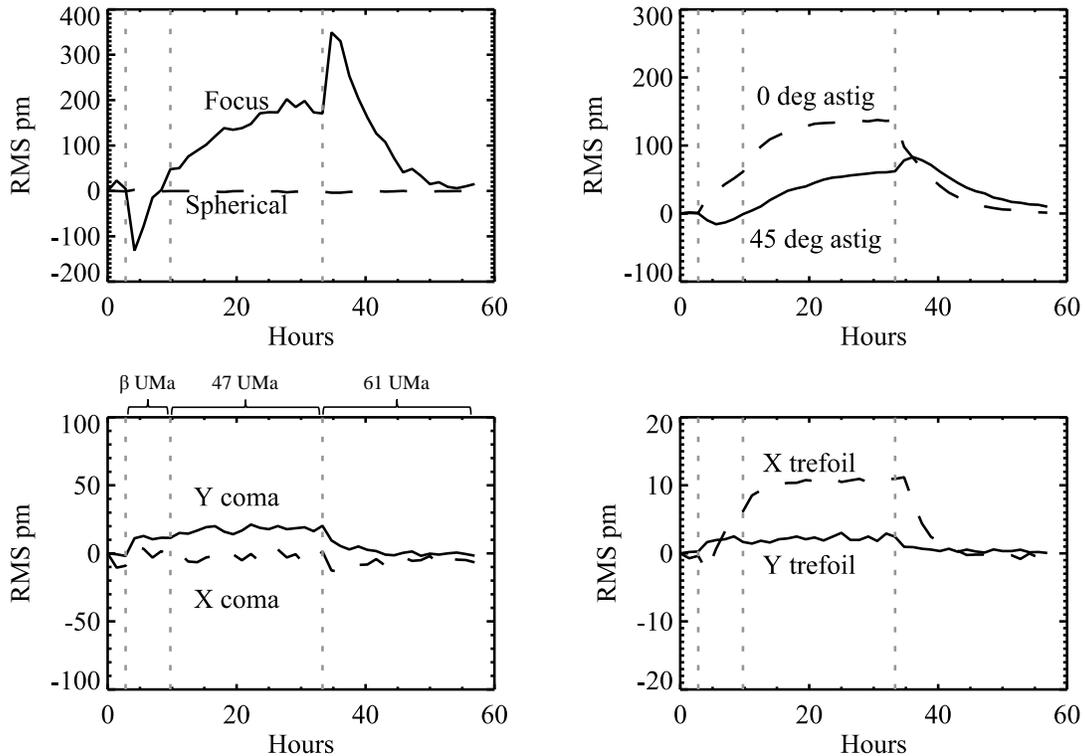

Fig. 37 - Low order wavefront variations over time expresses as Zernike polynomial coefficients (picometers RMS) at the plane of the FSM predicted by thermal and structural modelling for OS3 without the Earth (quasi-L2 orbit). The aberrations are referenced to time = 0 hours. The time spans spent slewing to and observing each star are indicated (prior to β UMa the system is pointed at 61 UMa).

*7.3 Time-dependent and low-order-corrected wavefront aberrations*

The low-order wavefront perturbation is propagated through system up to the LOWFS pickoff (the FPM in the case of the HLC or SPC, or the first Lyot stop in PIAACMC). From there it is fed into a separate model of the LOWFS where the wavefront disturbance is measured as a set of Zernike coefficients. The measurements include the effects of detector noise based on the brightness of the star and integration times. These Zernikes (up to spherical aberration, Z11 in Noll ordering) are then sent to a controller model that determines the motion of the focus adjustment mirror and the pattern to be added to the DM to correct them. The DM patterns, which include actuator gain errors, are added to the already-present dark hole solution in the PROPER model. The focus correction is added as a set of Zernikes in the PROPER model (the piston of the focus corrector mirror introduces some additional aberrations, which are determined from ray tracing and are included). The aberrated wavefront, which includes the correction for the previous time step, is then propagated through the full system to generate the low-order-corrected speckle field. The LOWFS/C system and its modelling are described in more detail by Shi et al. in this volume. Note that at the time of writing the effects of pointing and wavefront jitter have not been included in the time series models.

Fig. 37 shows the Zernike aberration changes predicted for OS3 without the Earth present (quasi-L2 orbit) from the combined thermal+structural+ray trace modelling computed at 5000 sec intervals. All of the aberrations change by less than a nanometer, but as the sensitivity results shown previously demonstrate, a few tens of picometers of an aberration can alter the speckle field significantly.

The thermal/structural models are run up to the FSM but not any further into the coronagraph. The two primary places where surface-motion-induced beam shearing of the mid-spatial-frequency aberrations can introduce any significant wavefront changes is between the primary/secondary and the two DMs in the HLC, as these optics have the largest wavefront errors in the system (~250 nm peak-to-valley for the DMs); the other optics have intentionally been specified to be super-smooth precisely to avoid beam walk effects from pointing offsets and optic motions and to reduce phase-induced amplitude errors. The transverse motion of the primary-to-secondary in the OS3 scenario is very low (maximum beam shear on the order of ~0.0001%). With the known aberrations on the primary and secondary, the contrast change is negligible. The thermal responses of the two DMs and what beam walk that might induce between them due to physical displacements - they have by far the largest wavefront aberrations when the HLC default settings are applied - are currently unkown. Small mid-spatial frequency aberration changes will affect the coronagraphs similarly; to first order, they bypass the focal plane mask because their resulting speckles occur further out in the field, unlike low-order aberrations that are strongly dependent on each coronagraph due to their interactions with the focal plane masks.

*7.4 Speckle field generation*

To illustrate the time-dependent effects in the dark hole, fields without and with DM LOWFC (but always with focus correction) were generated using the OS3 aberration curves for both the revised HLC and PIAACMC systems, as shown in the animated sequence in Fig. 38. Because a PIAACMC-specific LOWFS model was not available when this series was generated, the HLC LOWFS measurements and derived DM low-order correction patterns were used for the PIAACMC sequence. No jitter was included. The monochromatic fields comprising the broadband ones were weighted by the relative fluxes of the stars in each sub-band as computed from the appropriate model stellar spectra. Notice that the PIAACMC field without DM LOWFC pulses in brightness in relation to the change in astigmatism.

To more quantitatively demonstrate the variations in these speckle fields over time, for each time series the RMS difference between the field at each time step and the mean of all of the fields in that series was measured in an annulus between $r$ = 3 - 5 $\lambda_c$/D for HLC and PIAACMC, with and without DM LOWFC. As shown in Fig. 39, the HLC field RMS contrast variations are generally $<2 \times 10^{-10}$ even without DM LOWFC and about 50% lower with it. PIAACMC has $>10^{-9}$ changes without DM LOWFC, but with it the variations are reduced by at least an order of magnitude and are similar to HLC without it.

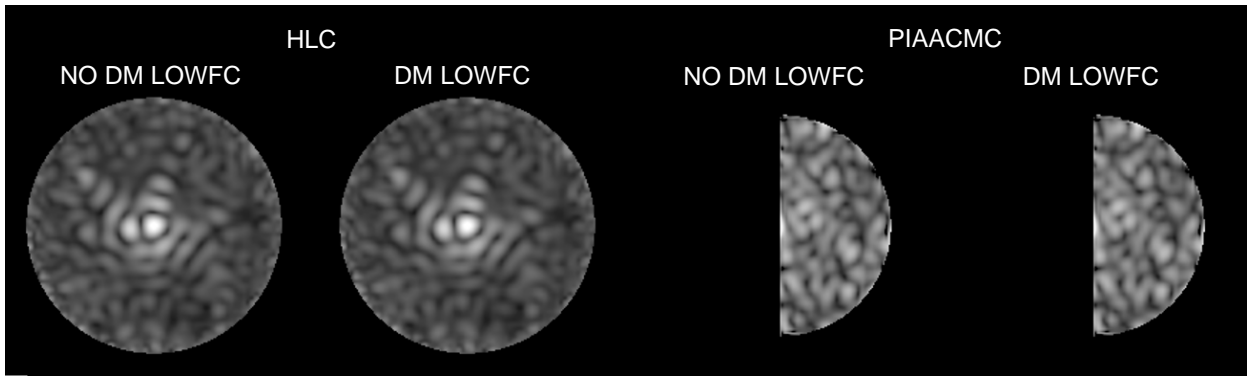

Fig. 38 - Animated sequence showing all frames from the OS3 (no Earth) time series simulations (5000 sec per frame). The revised HLC and PIAACMC fields without and with DM LOWFC are shown. The moving bar at the bottom denotes the relative place in the time series of the current frame. The images are shown between the contrast range of $10^{-10}$ - $10^{-8}$ with an intensity stretch of *contrast*$^{1/4}$ (MP4, 0.6 MB).

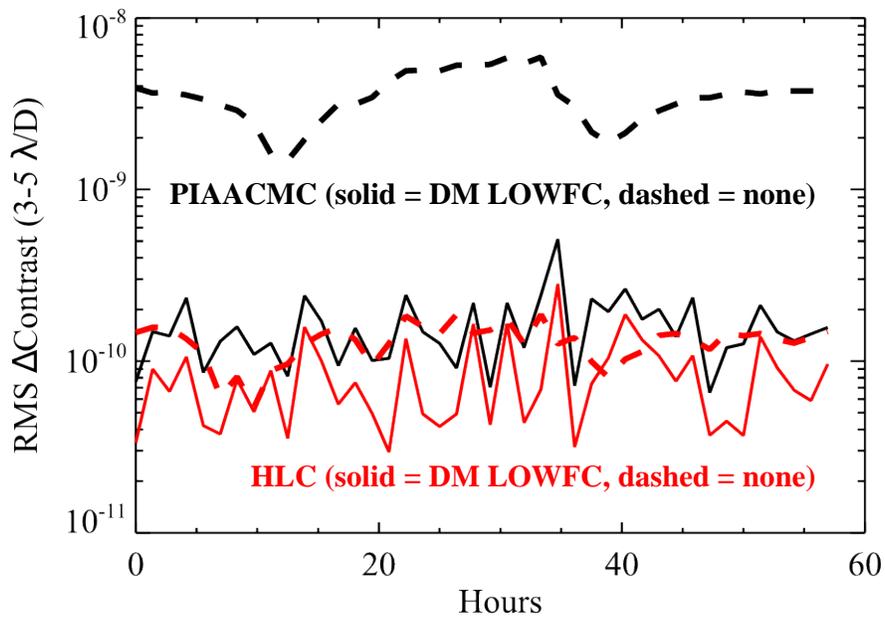

Fig. 39 - Plots of the RMS contrast variation of the difference between the OS3 field image computed at each time step and the mean of all of the images of the three stars in the time series, measured between $r = 3 - 5$ $\lambda_c/D$. Separate series are shown for HLC (red) and PIAACMC (black), both with (solid lines) and without (dashed lines) DM low order wavefront correction. These are the same fields shown in Fig. 38.

## 8 Post-processing

As has been demonstrated here, even with an optimized coronagraph and low and high order wavefront control, the mean contrast of the instrumental light in the dark hole field is typically equal to or greater than the contrast of the planets of interest ($10^{-8}$ - $10^{-9}$). To distinguish planets from speckles, which both look alike to some degree, some form of image processing is required. The most common method, both in space on the Hubble Space Telescope[39] and on ground-based telescopes, is subtracting an image or images of a reference star (sometimes the reference star is the same star observed at a different orientation). If the instrumentally-created speckles (and hence wavefront errors) remain perfectly stable during and between the science and reference star observations, then they will subtract out completely, down to the shot noise associated with those speckles. If the wavefront is unstable, the speckles will vary in time and lead to subtraction residuals that may either mimic planets or hide them. The ultimate performance of the coronagraph is therefore dependent on the combination of the entire optical system's wavefront error stability and that coronagraph's aberration sensitivities.

As a first step in the evaluation of the contrast performance after post-processing, three planets (Fig. 40) of various contrasts and separations were added to the 47 UMa frames of the OS3 no-Earth time series, with and without DM LOWFC. The frames corresponding to each star were then averaged together. The β UMa and 61 UMa images were separately subtracted from the 47 UMa image in the simplest form of classical PSF subtraction. Despite 61 UMa being a much closer spectral match to 47 UMa (both are G types), it provided no lower residuals than the β UMa (A type) subtraction (Fig. 41 and 42). The chromatic and time-dependent variations of the speckle fields are larger than the speckle mismatches caused by differences in stellar color. This is an important finding since close spectral matches to solar-type or lower (F, G, K) science target stars are likely to be as faint as those stars (V = 5 - 6, on average). Being able to use a much brighter star for the reference field, even with a considerably different spectral type, offers substantial savings in observing time to get the same signal-to-noise ratio exposures.

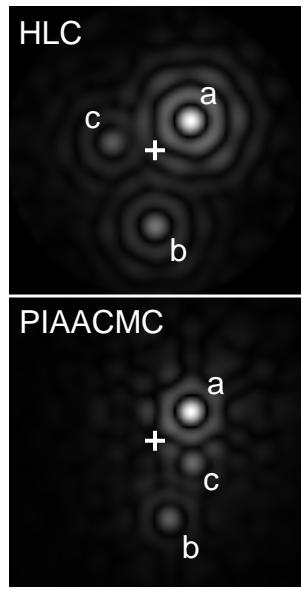

Fig. 40 - Test input scenes ($\lambda$ = 523 - 578, $\lambda_c$ = 550 nm) for the HLC and PIAACMC (including the appropriate PSFs for each field position) consisting of three planets with contrasts and apparent separations from the star of (a) $6 \times 10^{-9}$ at 3.1 $\lambda_c/D$, (b) $8 \times 10^{-10}$ at 5.3 $\lambda_c/D$, and (c) $9 \times 10^{-10}$ at 2.9 $\lambda_c/D$.

Fig. 41 shows that without DM LOWFC two of the three planets can be reliably detected in the HLC, but it is needed to see the third planet, which is closest to the star. The higher aberration sensitivity of PIAACMC coupled with no DM LOWFC results in perhaps only the brightest planet being reliably seen, and even then it is situated on high residuals and would possibly produce suspect photometry. Observing at multiple roll orientations of the telescope would likely allow for better discrimination between the residual speckles and planets. With DM LOWFC all three are clearly evident.

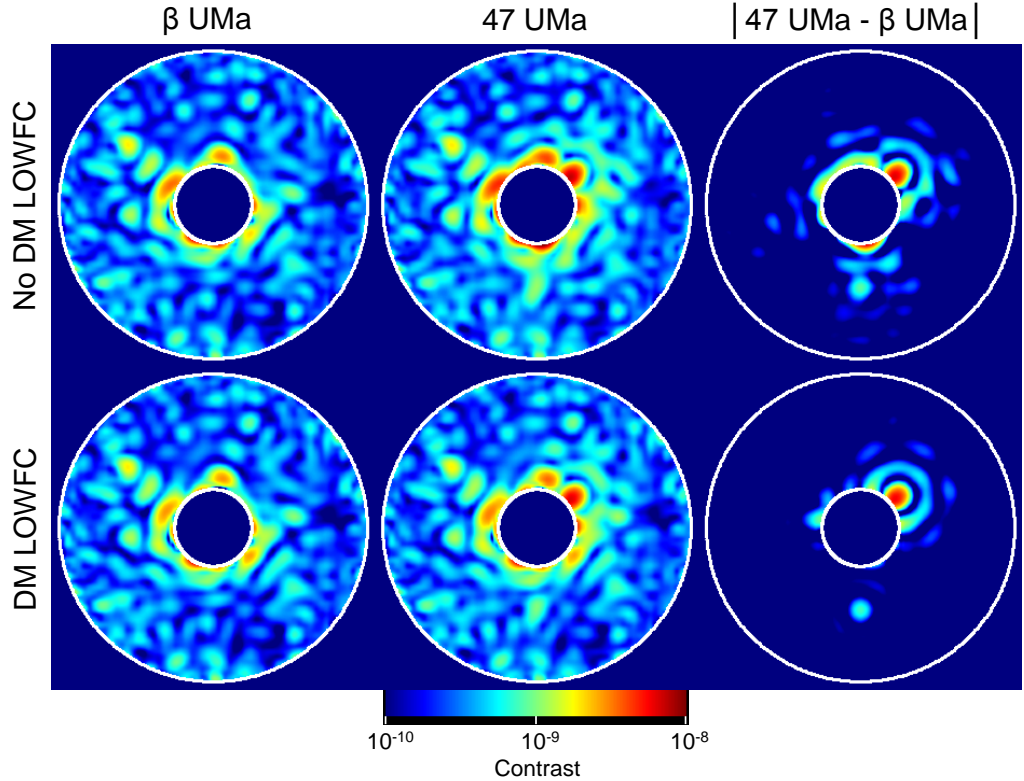

Fig. 41 - Averaged OS3 revised HLC dark hole fields ($\lambda$ = 523 - 578, $\lambda_c$ = 550 nm, single polarization) without detector noise for the bright star β UMa and science target 47 UMa (planets included) without and with low order wavefront control using the DM (focus correction is applied in all cases). The absolute differences between the two stars' fields is shown, revealing the planets. Without DM LOWFC only two planets can be seen, but with it all three can be identified. The circles are $r$ = 2.5 & 9.0 $\lambda_c$/D.

The raw (unsubtracted) and residual speckle noise, described in terms of the RMS over azimuth, are shown in Figs. 43 and 44. The improvement in the speckle noise levels at $r$ = 3 $\lambda_c$/D in the HLC is ~3× without DM LOWFC and ~22× with it. PIAACMC, with its higher aberration sensitivity, has essentially no improvement after subtraction without DM LOWFC but is ~18× better with it. These simulations show that the HLC can be scientifically productive without DM LOWFC, except near the IWA for ~$10^{-9}$ contrast sources. PIAACMC clearly requires DM LOWFC, and with it both coronagraphs achieve $10^{-10}$ or better speckle noise levels and allowing $10^{-9}$ contrast planets to be reliably seen (ignoring detector noise effects such as shot and read noise or dark current).

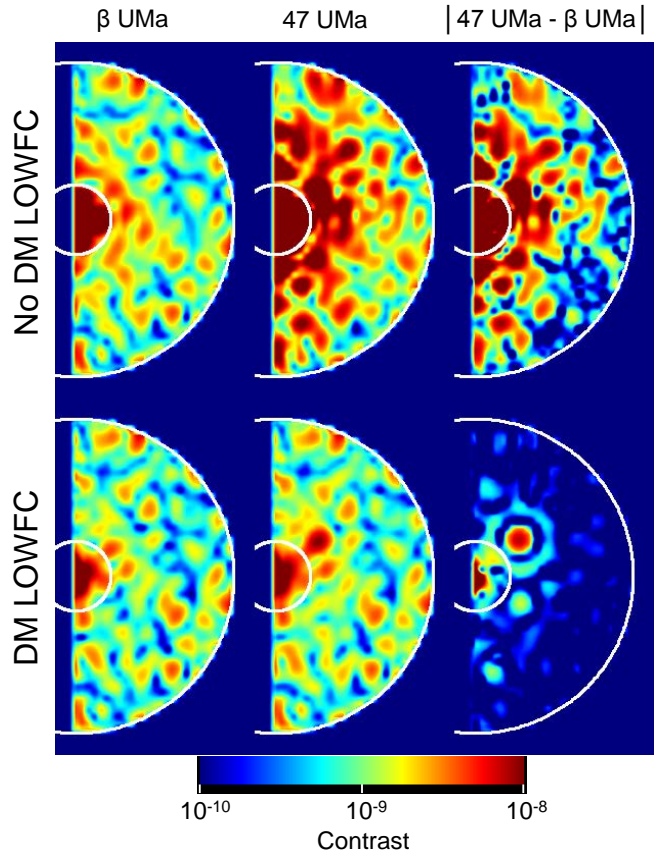

Fig. 42 - As with Fig. 41, but for PIAACMC (single polarization). The high sensitivity of PIAACMC to astigmatism causes the large variations in the fields between β UMa and 47 UMa when the DM LOWFC is not used. The circles are $r$ = 2.0 & 9.0 $\lambda_c$/D.

The classical PSF subtraction used here is the simplest way of reducing the raw speckle level, but more advanced and optimal methods exist. For example, the KLIP (Karhunen Loève Image Projection) algorithm[40] utilizes Principal Component Analysis to combine a series of science and reference field images to solve for the sky scene via projection onto eigenimages. To evaluate the application of KLIP and other techniques, the post-processing research group at the Space Telescope Science Institute[41] is analyzing these simulated datasets.

It should be noted that the results shown here are incomplete, and there are other effects that can alter the speckle stability over time. Chief among these are variations in pointing jitter. If the jitter is different between the science and reference star observations, either in amount or dominant direction of error, then differences in the speckle pattern will be introduced. This also includes differences between the diameters of the science and reference stars. Likewise, vibration-induced wavefront jitter may also vary over time (both effects are sensitive to the spacecraft reaction wheel speeds and configuration). These will be investigated in the future, along with the impact of differences in stellar diameters.

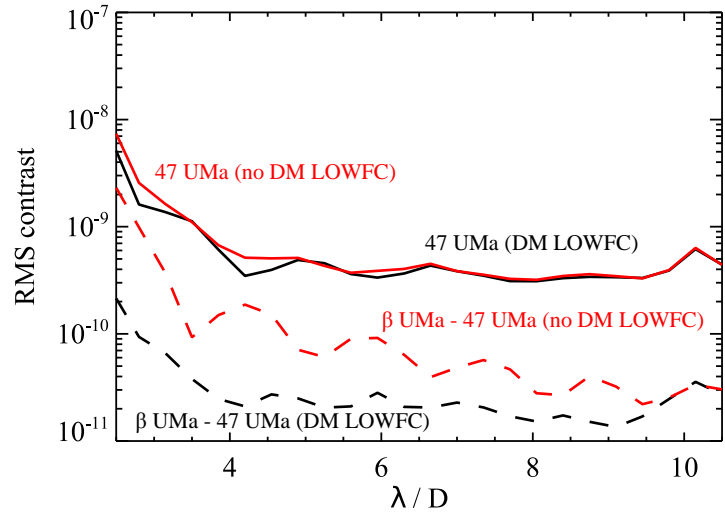

Fig. 43 - The revised HLC ($\lambda$ = 523 - 578, $\lambda_c$ = 550 nm, single polarization) OS3 time-averaged fields RMS contrasts measured over azimuth and plotted versus field radius presented as an indication of the speckle background noise against which a planet must be detected. The solid lines show the azimuthal RMS in the raw 47 UMa (planets omitted) field without (red) and with (black) DM LOWFC. The dashed lines show the azimuthal RMS of the difference between the β UMa and 47 UMa fields, representing the residual noise when classical PSF subtraction is applied.

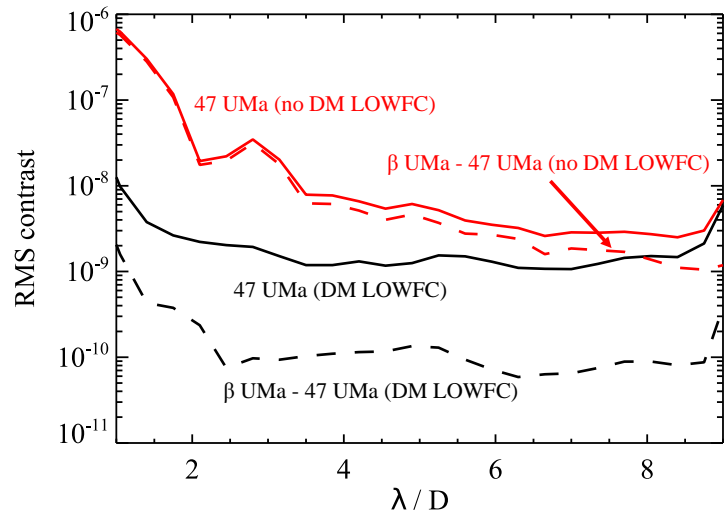

Fig. 44 - Same as Fig. 43 but for the revised PIAACMC (single polarization).

# 9 Conclusions

A summary of the coronagraphic performances are provided in Table 8. The results from modelling the three proposed WFIRST-AFTA planet imaging/characterization coronagraphs show:

(1) All three are capable of imaging Jupiter-contrast ($10^{-9}$) planets to within 2 - 3 λ/D of the star under the circumstances simulated here, subject to the amount of jitter present.

(2) Coronagraph diffraction suppression and telescope polarization typically set the static contrast limits.

(3) Wavefront stability sets the ultimate detection limits.

(4) The HLC provides the best contrasts for $r > 3$ λ/D over all pointing jitter and wavefront variations, and does not require but highly benefits from DM LOWFC.

(5) PIAACMC provides the highest effective throughput and smallest IWA, providing the best science return under ideal conditions, but it requires low order wavefront control using the DM to maintain wavefront stability.

(6) HLC and PIAACMC require polarization filtering to provide optimal contrast in the imaging channel.

(7) The SPC is well matched for use with the IFS due to its broad bandpass and relative insensitivity to polarization-induced aberrations.

(8) The reference star used in post-processing to remove the background speckles from the science image does not have to be very similar in spectral type as the speckle chromaticity is dominated by instrumental effects.

**Table 8. Summary of Revised Planet Imaging/Detection WFIRST-AFTA Coronagraphs**

| Coronagraph | Bandpass (nm) | Field Radius ($\lambda_c$/D) | Field Azimuth | IWA ($\lambda_c$/D) | PSF Core Throughput[c] | Mean Contrast[d] $r = 3 - 5$ $\lambda_c$/D 0.4 mas | 1.6 mas |
|---|---|---|---|---|---|---|---|
| HLC | 523 - 578 | 2.6 - 10.5[a] | 360° | 3.0 | 4.3% | $5 \times 10^{-10}$ | $2 \times 10^{-9}$ |
| SPC (IFS) | 728 - 872 | 2.5 - 9.0 | $2 \times 65°$ | 2.8 | 3.7% | $2 \times 10^{-9}$ | $4 \times 10^{-9}$ |
| PIAACMC | 523 - 578 | 1.2 - 9.0[b] | 180° | 1.3 | 14.0% | $5 \times 10^{-9}$ | $1 \times 10^{-7}$ |

[a] Inner radius set by occulter, outer by DM control

[b] Field radii set by wavefront control and inner radius transmission

[c] Unpolarized

[d] Single polarization (HLC, PIAACMC), dual polarization (SPC)

Some of the results presented here, notably the post-processing, are early attempts to evaluate the performance limits of the coronagraph under realistic conditions. A great deal more work is required to demonstrate the viability of the system designs, including:

(1) Incorporating wavefront sensing using DM probing patterns, including in the presence of jitter.

(2) Contrast evaluation and wavefront probing using finite area detector pixels, plus detector noise.

(3) Performing wavefront control in a limited number of finite bandwidth filters sampling the bandpass of the science filter.
(4) Evaluating the impact of uncorrectable high-temporal-frequency wavefront aberration changes due to vibration of optics by the spacecraft reaction wheels.
(5) Assessing the utility of advanced post-processing techniques (e.g., KLIP), especially in the presence of detector noise.
(6) Evaluating the HLC and PIAACMC at other bandpasses (this requires new FPMs and DM patterns).
(7) Validating models against testbed results.

These topics will all be addressed within the next couple of years as the WFIRST-AFTA coronagraphs transition from technology demonstrations to flight ready designs.

*Acknowledgments*

This research was carried out at the Jet Propulsion Laboratory, California Institute of Technology, under a contract with the National Aeronautics and Space Administration. The authors thank Brian Kern, Wes Traub, Dwight Moody, Dimitri Mawet, Erkin Sidick, Hanying Zhou, Stuart Shaklan, and Gary Gutt for discussions relating to these results.

35. O. Guyon, "Phase-induced amplitude apodization of telescope pupils for extrasolar terrestrial planet imaging," *Astron. and Astrophys.* **404**, 379-387 (2003).
36. O. Guyon, et al. (this volume) (2015).
37. L. Pueyo, S. Shaklan, A. Give'on, and J. Krist, "Numerical propagator through PIAA optics," Proc. SPIE. 7440, 74400E (2009).
38. R. Soummer, L. Pueyo, A. Sivaramakrishnan, and R. Vanderbei, "Fast computation of Lyot-style coronagraph propagation," *Optics Express.* **15**, 15935 (2007).
39. J. Krist, "High-contrast imaging with the Hubble Space Telescope: performance and lessons learned," Proc. SPIE. 5487, 1284 (2004).
40. R. Soummer, L. Pueyo, and J. Larkin, "Detection and Characterization of Exoplanets and Disks Using Projections on Karhunen-Loève Eigenimages," *Astrophys. J. Letters.* **755**, L28 (2012).
41. Debes et al., (in prep) (2015).
50